\documentclass[conference]{IEEEtran}
\IEEEoverridecommandlockouts
\usepackage{cite}
\usepackage{amsmath,amsthm,amssymb,amsfonts,mathtools}
\usepackage{algorithmic}
\usepackage{graphicx}
\usepackage{textcomp}
\usepackage{xcolor}

\newtheorem{remark}{Remark}
\mathtoolsset{showonlyrefs=true}

\def\BibTeX{{\rm B\kern-.05em{\sc i\kern-.025em b}\kern-.08em
    T\kern-.1667em\lower.7ex\hbox{E}\kern-.125emX}}
    
\begin{document}
\thispagestyle{empty}
Submitted to: the 2027 IEEE International Conference on Robotics \(\&\) Automation (ICRA).

Copyright may be transferred without notice, after which this version may no longer be accessible.
\newpage

\title{PDE-Constrained MPC of Motility-Induced Phase Separation in Robotic Swarms: The Role of Model Fidelity and Local Information}
\author{Longchen Niu and Gennaro Notomista%
\thanks{The authors are with the Department of Electrical and Computer Engineering, University of Waterloo, Waterloo, ON, Canada {\tt\footnotesize {l3niu,gennaro.notomista}@uwaterloo.ca}}%
}
\maketitle

\begin{abstract}
Motility-induced phase separation enables swarm aggregation to be regulated without prescribing a target spatial distribution. This paper develops a partial differential equation-constrained model predictive control framework for a vibration-driven robotic swarm. Three continuum models retaining progressively less orientation information are considered to study the trade-offs among model fidelity, information requirements, and computational cost. The framework is evaluated under centralized and decentralized settings, with focus on how limited sensing and local-to-global phase-separation estimation affect the performance of predictive control.
\end{abstract}

\section{Introduction}
In recent years, robotic swarms consisting of many simple units have gained traction over traditional, complex single-robot solutions for applications where robustness, resilience, redundancy, and scalability are of utmost importance. These applications include environmental monitoring, precision agriculture, search and surveillance, and biomedical micro-manipulations \cite{environment_monitor,agriculture,search_app,medical}. In these scenarios, robots are required to alternate between dispersed formations, which encourage exploration and information gathering, and aggregated configurations, which foster cooperation, communication, manipulation, and information aggregation. Although consensus methods can drive robots toward a common state, they do not directly regulate macroscopic properties such as swarm shape, cluster size, or degrees of aggregation. For this reason, density-based approaches have been developed so that useful swarm behaviors can emerge from simple individual actions \cite{density_shape,density_cluster,MIPS_index}.

A physical mechanism for such aggregation is provided by the so-called Motility-Induced Phase Separation (MIPS), in which self-propelled particles separate into dilute and dense phases with only repulsive forces at play \cite{mips}. Collisions reduce particle speed, which makes future collisions more likely to happen at lower speed, forming a positive feedback loop. The authors of \cite{PDE_Math_derive} showed that active Brownian particles and run-and-tumble particles admit equivalent coarse-grained formulations when their motility depends on density and not orientation. Their formulation provides the mathematical framework underlying the models considered in this paper: (i) a Fokker-Planck (FP) Partial Differential Equation (PDE) retaining the full position and orientation information, (ii) an angular-momentum-focused model retaining density and polarization, and (iii) a density-only diffusion equation, under long-time and long-length scale assumptions. These models progressively reduce the required orientation measurements, allowing us to study how this simplification affects prediction and control behavior.
More recently, the work in \cite{micro_to_macro} systematically derived four different microscope repulsion models to macroscopic PDEs and examined their stability and MIPS behaviors using simulations.

Physical experiments and numerical simulations have further established the mechanisms and broadened the scope of MIPS.  
The authors in \cite{Janas_exp} observed self-trapping, clustering, and phase separation in experiments with heat-driven Janus particles, and reversible phase separation has also been produced in photosynthetic microbes \cite{light_micro_exp}. More recent studies have extended the classical MIPS framework beyond unbiased Brownian particles. For example, chemotactic particles can produce phase separation through a mixture of chemotaxis and density-dependent motility \cite{chemotactic}. Similarly, the authors of \cite{MIPS_Traffic_OneD} studied the effects of activity and directional bias in one-dimensional traffic models. Collectively, these results demonstrate that phase separation can be regulated through different types of inputs, such as light, chemical fields, and directional bias, revealing a pathway for controlling MIPS as a collective behavior in engineered robotic swarms.

Active-matter principles have consequently begun to influence the design and control of robotic swarms. Using \(300\) micro-bristle robots on a shaker plate, the authors of \cite{MIPS_index} demonstrated collision-induced aggregation, introduced the MIPS index as a parameter-free measure of aggregation, and regulated the aggregation level by switching between vibration modes with a bang-bang type of controller. Field-driven active colloids have similarly been organized into reconfigurable non-equilibrium patterns using externally applied electric and magnetic fields \cite{field_driven}. For systems requiring individual actuation, vibration-driven robots (e.g., \cite{dmitriev2023swarmodroid,brushbot}) combine simple mechanical designs with onboard motors, computation, and wireless communication, allowing each robot’s motion to be controlled independently. At the swarm level, however, these approaches largely rely on prescribed commands or switching between predefined actuation modes. A more general feedback approach was proposed in \cite{RalPaper}, where a decentralized, online PDE-constrained controller used the FP model to steer the swarm density toward predefined spatial distributions, with experimental validation on brushbots. Because this method optimizes only the next predicted state, it does not capture nonlinear swarm evolution over longer horizons. To address this, \cite{brownian_MPC} embedded a Smoluchowski model of non-interacting active Brownian particles into an MPC framework to regulate density and orientational moments.
Nevertheless, these methods require a prescribed density, polarization, or flux field. When this field is unknown, or its exact shape matters less than the aggregation level, such as in exploration or environmental monitoring, the degree of phase separation provides a more natural, location-independent control objective. To the best of our knowledge, no existing approach directly regulates this quantity.
Therefore, in this work, we develop a PDE-constrained MPC framework that, leveraging the MIPS index, controls phase separation of a vibration-driven robotic swarm.

The main contributions of this work are: (i) a continuous formulation of the MIPS-index dynamics and a PDE-constrained MPC framework for tracking aggregation levels without prescribing a target spatial distribution; (ii) a unified comparison of three continuum predictors retaining progressively less orientation information; and (iii) a decentralized formulation that quantifies how detection radius and local-to-global MIPS estimation affect closed-loop performance.

\section{Mathematical Models}
\label{sec: math background}
MIPS is the separation of dense, slow-moving clusters and sparse, fast-moving regions. To quantify this phenomenon, the authors of \cite{MIPS_index} proposed a novel dimensionless MIPS index that evaluates the extent of this separation by measuring the variance of the density field over multiple spatial windows. 

In this section, we recap the definition of the MIPS index used to quantify swarm aggregation levels, followed by the microscopic kinematics of individual robots. Next, we report three macroscopic density models presented in \cite{PDE_Math_derive} as the framework to describe the swarm's continuum evolution. These models will be used in the next section to derive the MIPS index dynamics.

\subsection{MIPS Index}
Let \(\Omega \subset \mathbb{R}^2\) denote the 2D operation domain, with area \(A\), and position \(z_i\) for robot \(i\). Following \cite{MIPS_index}, the swarm is sampled at the lattice points of an \(l \times l\) grid. For each grid point \(r' \in \{0,1, \dots, l-1\}^2\), a square window with side length \(w\) is defined as \(\mathcal{W}(r',w)\). The resulting discrete coarse-grained density over the sampling window is:\(
[\lambda^{\mathrm d}_{w}(t)]{r'} =\frac{\left| {z_i(t): z_i(t)\in\mathcal{W}(r',w) }\right|}{|\mathcal{W}(r',w)|}.\)
To enable PDE-based prediction, we replace the discrete lattice measurement \([\lambda_w^{\mathrm d}(t)]_{r'}\) in \cite{MIPS_index} with a continuous window-averaged density \(\lambda_w(r',t)\) constructed from the macroscopic field \(\lambda(r,t)\) derived in Section~\ref{sec: density_model}.

For a continuous macroscopic density \(\lambda(r,t)\), its local average over the window \(\mathcal{W}(r',w)\) is
\(
\lambda_w(r',t) = \frac{1}{|\mathcal W(r',w)|}
\int_{\mathcal W(r',w)}
\lambda(r,t)\,dr,\)
with spatial variance:
\begin{equation}
\operatorname{Var}(\lambda_w(t))=\frac{1}{A}
\int_{\Omega} \left( \lambda_w(r',t) - \bar{\lambda}_w(t) \right)^2 dr',
\label{eq:continuous_var_lambda}
\end{equation}
where
\(\bar{\lambda}(t) = \bar{\lambda}_w(t) = \frac{1}{A} \int_\Omega \lambda(r,t)\,dr \)
is constant for all window sizes by mass conservation under reflective or periodic boundary conditions when constructing \(\lambda(r,t)\).

To ensure the resulting measurement is independent of window sizes, the density variance is therefore averaged over a set of finite window sizes in the interval \([w_{\min}, w_{\max}]\) as:
\(
I_A(t)=\frac{1}{w_{\max} - w_{\min}} \int_{w_{\min}}^{w_{\max}} \operatorname{Var}(\lambda_w) dw.\)
This averaging step offers an advantage over existing cluster-measuring techniques based on adjacency relations, density-based clustering, and sampling-based methods, as it reliably captures phase separation in a swarm without requiring a posteriori manual parameter tuning. Furthermore, to ensure the MIPS index is agnostic of swarm or environment sizes, the final measurement is normalized into \([0,1]\) by:
\begin{equation}
\text{MIPS index}
=
\frac{I_A(t)-I_{A,\min}}
{I_{A,\max}-I_{A,\min}},
\label{eq: mips_def}
\end{equation}
where \(I_{A,\max}\) and \(I_{A,\min}\) correspond to the maximum and minimum attainable values, representing a single tightly packed cluster and uniform distribution over the full domain. 

\subsection{Robot Motion Dynamics}
While the MIPS index above measures the swarm's instantaneous aggregation level, it does not explain how each robot's motion contributes to it. To this end, we introduce a microscopic motion model for individual robots in the swarm, which is used to derive the evolution of the macroscopic swarm density field used to calculate the MIPS index. 

The micro-bristle robots introduced in \cite{MIPS_index} serve as a prominent foundation for this, as they exhibit persistent forward motion with stochastic angular noise. Let $z_i(t)=[x_i(t),y_i(t)]^\top\in \Omega$ and $\theta_i(t)\in[0,2\pi)$ denote the position and orientation of robot $i$, respectively. Within the closed and bounded domain \(\Omega\), we define the boundary of the circular test field as an inward-sloping surface, characterized by the annular set \(\Delta\Omega \in \mathbb{R}^2\). Let \(\mathbb{N}_i\) denote the set of robots that robot \(i\) is colliding with, and define \(N_i = \frac{|\mathbb{N}_i|}{N_{\max}}\) as its normalized cardinality, with \(N_{\max}\) as the maximum number of allowed collisions by robot geometry. 

The individual kinematics of the robot can be represented by the following active Brownian particle model:
\begin{equation}
\begin{aligned}
\dot{x}_i&=\big(v_0 \cos\theta_i + \mu \sum_{j\in\mathbb{N}_i} f_{cx}(x_i-x_j) \big)s_{t,i}f_i,\\
\dot{y}_i &= \big(v_0 \sin\theta_i+\mu \sum_{j\in\mathbb{N}_i}
f_{cy}(y_i-y_j)
\big)s_{t,i}f_i,\\
d\theta_i&=\big(\sqrt{2D_r}\,\Delta W_\theta + s_{r,i} \big) f_i,
\end{aligned}
\label{eq: motion model}
\end{equation}
where \(\Delta W_\theta\) is the Wiener process increment representing Gaussian noise. The linear free speed \(v_0\) and angular diffusion coefficient \(D_r\) are determined by the controlled vibration frequency and amplitude. Inter-robot collision is characterized by the motility coefficient \(\mu\) in the inverse normal direction, defined by:
\(
f_{cx}(x_i-x_j)
=
\frac{x_i-x_j}
{\sqrt{(x_i-x_j)^2+(y_i-y_j)^2}}
,
f_{cy}(y_i-y_j)
=
\frac{y_i-y_j}
{\sqrt{(x_i-x_j)^2+(y_i-y_j)^2}}.\)
The friction effect \(f_i\) is characterized by the nonlinearity coefficient \(\beta\) and the maximum magnitude \(F\): \(
f_i =  1-\frac{F\beta N_i} {1-(1-\beta)N_i}.\)
Finally, let \(\phi_i=\tan^{-1}(y_i/x_i)\). The sloped edge is modeled by setting \(s_{t,i}=1-L\) and \(s_{r,i}=(1-L)(\pi/2)\sin(\theta_i-\phi_i)\) for \(z_i\in\Delta\Omega\), and \(s_{t,i}=s_{r,i}=1\) otherwise. Here, \(L\in(0,1)\) is a scaling factor, while \(s_{r,i}\) rotates robots toward the arena center to reduce boundary collisions.

\subsection{Density Model of Active Brownian Particles}
\label{sec: density_model}
Although the microscopic model describes individual robot trajectories, directly propagating the robot states becomes computationally challenging with hundreds or thousands of robots. A continuum model is therefore introduced to describe the evolution of the density state dynamics of the swarm. The density-based representation provides a common framework for deriving three different models using varying amounts of heading information from each robot in the swarm. 

Let $\psi(r,\theta,t)$ denote the position-orientation density, representing the probability of finding a robot at location \(r\) with heading \(\theta\) at time \(t\). Then, the spatial robot density can be obtained as: \(
\lambda(r,t) = \int_{0}^{2\pi} \psi(r,\theta,t)\,\mathrm{d}\theta.\)

\subsubsection{Fokker-Planck Model}
Persistent linear motion with stochastic angular diffusion implies that future spatial distribution cannot be precisely predicted from \(\lambda\) alone, without consideration of current robot headings. Therefore, we first introduce a FP model without eliminating the angular state. This full information model serves as the highest-fidelity model and as the foundation for the other two models.

The stochastic microscopic motion model in \eqref{eq: motion model} induces the FP equation \cite{PDE_Math_derive}:
\begin{equation}
\frac{\partial\psi}{\partial t} = 
-\nabla_r
[ v(\lambda)e(\theta)\psi ] + \nabla_\theta [D_r \nabla_\theta \psi].
\label{eq: fp_model}
\end{equation}
The first term describes transport along the robot heading with the density-based swim speed \(v(\lambda)\) depending only on positions, not heading \(e(\theta) = [\sin(\theta), \cos(\theta)]^T\).
The second term characterizes the effects of rotational diffusion. Equation \eqref{eq: fp_model} evolves the complete position-orientation distribution and consequently preserves all directional information as an extra state. Consequently, this is the highest fidelity model, with the trade-off of the highest computational demand.

\subsubsection{Density-Polarization Model}
Although the full FP equation utilizes the complete orientation information, evolving $\psi(r,\theta,t)$ requires an additional angular grid, increasing computation cost for large swarms. Conversely, reducing the model directly to the spatial density $\lambda(r,t)$ would completely discard the directional memory of the robots. This information is important over timescales comparable to $1/D_r$, since two swarms with the same density can evolve drastically differently if their robots have different headings.

A compromise, and an intermediate step, is obtained by decomposing \(\psi(r,\theta,t)\) into its zeroth, first, second, and higher-order angular harmonic components \cite{PDE_Math_derive}:
\(
\psi = \lambda + e(\theta) p + Q + \Theta[\theta].
\)
Retaining only the first harmonic component as an approximation to the full decomposition for computational speed and numerical stability, the resulting polarization field,
\(
p=\int_{0}^{2\pi}e(\theta)\psi(r,\theta,t) \,\mathrm{d}\theta, 
\)
represents the net local direction of motion and directly determines the active density flux, $v(\lambda)p$. The density-polarization model therefore eliminates the explicit \(\theta\) state while preserving the first moment of the robots' directional bias. 

Integrating \eqref{eq: fp_model} over $e(\theta)$ gives the exact conservation law
\begin{equation}
\frac{\partial\lambda}{\partial t} = -\frac{1}{2} \nabla \left(v(\lambda)p\right), \quad
\frac{\partial p}{\partial t} = -\nabla\left(v(\lambda)\lambda\right) -D_r p,
\label{eq: polarization model}
\end{equation}
up to the first angular harmonic component. 
Equation \eqref{eq: polarization model} shows that spatial density is transported by the polarization, which is generated by gradients in $v(\lambda)\lambda$ and decays through rotational diffusion.
This model reduces the dimensionality of the full FP formulation while retaining the dominant first-order transient directional response of the robots.

\subsubsection{Density-Flux Model}
Although the density-polarization model eliminates the explicit angular state, it still requires the polarization field $p$ to be estimated and evolved with \(\lambda\) as a coupled PDE system. This is undesirable for density-based feedback, since the MIPS index directly requires $\lambda$, but not any measurement of \(\theta\) and, therefore, $p$. Furthermore, evolving the additional density field increases the computational cost and threatens the numerical stability of model predictive control frameworks. Therefore, a density-only model is developed under assumptions of time and length scales specified in \cite{PDE_Math_derive}.

When the polarization relaxes sufficiently quickly relative to the evolution of $\lambda$, it can be approximated by its quasi-steady value rather than propagated as an independent state. Eliminating $p$ in this manner gives a closed conservation law:
\begin{equation}
\partial_t\lambda=-\nabla\ J(\lambda), \quad J = - \frac{v(\lambda)}{2 D_r} \nabla (v(\lambda) \lambda),
\label{eq: dlambda dt}
\end{equation}
where the density flux depends only on $\lambda$, $v(\lambda)$, and $D_r$. By reducing the model to a single scalar field, it provides the simplest connection between the vibration-dependent motility parameters and the evolution of the swarm density, and therefore the MIPS index. This computational and analytical simplification is possible when the polarization relaxes sufficiently quickly relative to changes in the density field.
In particular, the approximation requires time and length scales satisfying
\begin{equation}
t > \tau_r=\frac{1}{D_r},
\qquad
L> \ell_p=\frac{v_0}{D_r},
\label{eq scale require}
\end{equation}
where $\tau_r$ and $\ell_p$ are the orientational relaxation time and persistence length, respectively.
At these scales, $\partial_t p \approx 0$ and higher-order terms are neglected in \eqref{eq: polarization model}, giving the quasi-steady approximation
\(
p\approx
-\frac{1}{2D_r}
\nabla\left(v(\lambda)\lambda\right).
\)
Applying the chain rule and substituting into \eqref{eq: dlambda dt} yields a density-only conservation law:
\begin{equation}
\frac{\partial\lambda}{\partial t}
=
\nabla (D_{\mathrm{eff}}(\lambda)\nabla\lambda),  \quad D_{\mathrm{eff}}(\lambda)
=
\frac{v(\lambda)}{2D_r}
\left(
v(\lambda)
+\lambda\frac{\mathrm{d}v}{\mathrm{d}\lambda}
\right).
\end{equation}
Mathematically, a pure negative diffusion PDE is ill-posed forward in time; therefore, the MPC trying to simulate it forward to predict the future state of the swarm will encounter numerical instability. To make the prediction numerically stable, one could add a regularizer term, which limits how sharp the boundaries of the clusters can get and turns the active matter PDE into a generalized Cahn-Hilliard equation:
	\begin{equation}
		\frac{\partial \lambda}{\partial t} = \nabla \left( D_{\text{eff}}(\lambda) \nabla \lambda \right) - \gamma \nabla^4 \lambda,
        \label{eq: full density only model}
	\end{equation}
where $\gamma > 0$ is a small penalization constant for large spatial gradients. This ensures that the forward predictions remain stable even during extreme aggregation phases, characterized by a large MIPS index.

Equation \eqref{eq: full density only model} provides the connection between density-dependent slowdown and phase separation. In particular, the effective diffusivity becomes negative when
\(
v(\lambda)
+\lambda\frac{\mathrm{d}v}{\mathrm{d}\lambda}<0.\)
Small density perturbations then grow rather than diffuse, producing dense and dilute regions. However, \eqref{eq: full density only model} is valid only under the scale assumptions in \eqref{eq scale require}. The simulation parameters are selected to satisfy these assumptions, enabling the model to be used within the MPC framework.
The density-polarization and full FP models are retained to determine whether explicitly modeling orientation improves prediction and control.

\section{MIPS Dynamics and Control}
\label{sec: mips dynamics control}
We are now ready to derive the time evolution of the MIPS index, which results in a PDE that is based on the evolution of the spatial density of the swarm. We then formulate a centralized MPC based on the MIPS index PDE dynamics, aiming to predict and track any, possibly time-varying, target aggregation level. Finally, for individually actuated robots, we develop a decentralized formulation based on local observations, enabling us to study how information availability affects prediction and closed-loop performance.

%A decentralized formulation of such a controller is also proposed for applications where individual robots can be controlled independently. 

\subsection{Dynamics of the MIPS Index}
The preceding continuum models provide different approximations of $\partial_t\lambda$, but to track desired phase separations, the control objective needs to be expressed in terms of the scalar MIPS index \eqref{eq: mips_def} rather than the density field itself. To connect the continuum dynamics to this objective, in this section, we differentiate the MIPS index with respect to time. The resulting PDE produces a common expression through which the full FP \eqref{eq: fp_model}, density-polarization \eqref{eq: polarization model}, and density-only models \eqref{eq: full density only model} can be used to predict the instantaneous rate of aggregation.

We start by taking the time derivative of the MIPS index: 
\begin{equation}
\begin{aligned}
    &\frac{d}{dt} \text{MIPS}(t) = \frac{1}{I_{A,\max} - I_{A,\min}} \frac{d}{dt} I_A(t)\\
    &= \frac{1}{I_{A,\max} - I_{A,\min}} \frac{1}{w_{\max} - w_{\min}} \int_{w_{\min}}^{w_{\max}} \frac{d}{dt} \text{Var}(\lambda_w(t)) dw.
\end{aligned}
\label{eq:d_dt_MIPS}
\end{equation}
Since the total number of robots is conserved, the mean density \(\bar{\lambda}\) is constant for all window sizes. The time derivative of eq.~\eqref{eq:continuous_var_lambda} is then solved by chain rule: 
\begin{equation}
    \frac{d}{dt} \text{Var}(\lambda_w(t)) = \frac{2}{A} \int_\Omega \left( \lambda_w(r',t) - \bar{\lambda} \right) \frac{\partial \lambda_w(r',t)}{\partial t} dr',
\end{equation}
where
\(
\frac{\partial \lambda_w(r',t)}{\partial t} = \frac{1}{w^2} \int_{\mathcal{W}(r',w)} \frac{\partial \lambda(r,t)}{\partial t} dr.
\)
The swarm density evolution, \(\frac{\partial \lambda(r,t)}{\partial t}\), can be implemented with any of the three continuum models:  \eqref{eq: fp_model}, \eqref{eq: polarization model}, or \eqref{eq: full density only model}. 

\subsection{Controller Formulation}
\subsubsection{Centralized MPC}
Let \(M(\lambda(r,t))\) denote the MIPS index introduced above. At each control update \(t_0\), the centralized controller has access to the positions and headings of all robots and constructs a global continuum density state estimate over \(\Omega\). A single control input
\(
u(t) = \begin{bmatrix}
f(t) \quad a(t)
\end{bmatrix}^T
\)
is applied uniformly to the entire swarm in the form of a global vibration frequency and amplitude.
The control input affects the density dynamics through the data-driven
relationships
\begin{equation}
v(\lambda) = v_0(u) \ (1 - \xi_1 \lambda - \xi_2 \lambda^2 - \xi_3 \lambda^3),\quad D_r=D_r(u),
\end{equation}
which map the vibration parameters to the density-dependent swim speed
and angular diffusion coefficient by fitting the experimental data collected in \cite{MIPS_index}. 
The objective is to track a prescribed MIPS-index trajectory \(M^*(t)\) over the prediction horizon \(T\), the resulting MPC is:
\begin{equation}
    \begin{aligned}
        \min_{{u}(t)} & \int_{t_0}^{t_0+T} \left( \text{MIPS}(\lambda({r},t)) - M^*(t) \right)^2 dt \\
        &+ \int_{t_0}^{t_0+T} {u}(t)^T {R} {u}(t) dt\\
        s.t. \quad & \text{MIPS index dynamics: } {\eqref{eq:d_dt_MIPS}} \\
        & \text{Continuum density dynamics: }  \eqref{eq: fp_model} \, or \,  \eqref{eq: polarization model} \, or \,  \eqref{eq: full density only model} \\
        & \lambda({r},t_0) = \mathcal{G}(z(t_0)),
    \end{aligned}
    \label{eq: centralized MPC}
\end{equation}
where $R$ is a positive definite weight matrix balancing target tracking and energy conservation, $\mathcal{G}$ is a density constructor that utilizes a Gaussian kernel to initialize corresponding density functions from available robot positions, \(z = [z_0^T, z_1^T, \dots, z_N]^T \), and headings, \(\theta = [\theta_0, \theta_1, \dots, \theta_N]\), for \(N\) robots in the swarm.
At each update, the optimization uses the target trajectory
\(M^*([t_0, t_0+T])\) over the prediction horizon. At the next timestep, the global continuum state is reconstructed, and the optimization is repeated.

This formulation is directly applicable to passive or globally
actuated swarms, including micro-bristle robots on a shaker plate
\cite{MIPS_index}, microorganisms under uniform illumination
\cite{light_micro_exp}, and active colloids driven by external electric
or magnetic fields \cite{field_driven}. Because the shared input
\(u\in\mathbb R^2\) does not grow with the swarm size, the primary threat to feasibility is communication delay and operating domain size.
For a swarm with agents capable of individual actuation, however, a centralized formulation would instead optimize for
\(
u = [u_1^T,  \cdots,  u_N^T]^T \in\mathbb R^{2N},
\)
while also collecting the state of every robot. The rapid growth in decision variables and communication load motivates the following decentralized
formulation.

\subsubsection{Decentralized MPC}
A shared input is suitable for globally actuated swarms, whereas individual actuation would require \(2N\) decision variables and global state collection, motivating the decentralized formulation below.

%As discussed above, extending the centralized MPC to a swarm with individual actuation would drastically increase the associated communication and optimization requirements as the swarm size grows. We therefore formulate a decentralized MPC in which each robot constructs a local continuum density state and independently optimizes its own actuation using only locally available information.

In applications without global communication, robots typically only have sensing capabilities of the environment around them within some detection radius \(D\). We define a distance-based sensing graph, \(\Delta\)-disk graph \cite{mesbahi2010graph}, as the neighborhood set of robot \(i\): \(\mathcal{N}_i = \left\{ j \in N\, \middle| \, \| {z}_j - {z}_i \| \leq D \right\}\), including \(i\in\mathcal N_i\). The corresponding local observation domain is
\(
\Omega_i(t) = \Omega\cap B_D(z_i(t)).
\)
Each robot obtains the positions of its neighbors and, for the
orientation-dependent predictors, their headings within this domain.

\begin{remark}
\label{remark:Collsion_based_decentral}
Recall that \(\mathbb N_i(t)\) denotes the collision-neighbor set of robot \(i\). If \(D\) is at least the maximum inter-robot collision distance, then \(\mathbb N_i(t)\subseteq\mathcal N_i(t).\) 
Consequently, the same decentralized formulation remains applicable
when sensing is restricted to collision detection by selecting \(D\) equal to the collision distance. The optimization structure is
unchanged, although the local-to-global target calibration and resulting
control performance remain dependent on \(D\).
\end{remark}

We develop the following decentralized MPC tailored for independently controllable swarms:
\begin{equation}
    \begin{aligned}
        \min_{{u}_i(t)} & \int_{t_0}^{t_0+T} \left( \text{MIPS}_D(\lambda_{i}({r},t)) - M^*_{D}(t) \right)^2 dt \\
        & + \int_{t_0}^{t_0+T} {u}_i(t)^T {R} {u}_i(t) dt\\
        s.t. \quad & \text{Local MIPS index dynamics from: } {\eqref{eq:d_dt_MIPS}} \\
        & \text{Local density dynamics from: }  \eqref{eq: fp_model} \, or \,  \eqref{eq: polarization model} \, or \,  \eqref{eq: full density only model} \\
        & \lambda_{i}({r},t_0) = \mathcal{G}(z_{\mathcal{N}_i}(t_0)),
    \end{aligned}
    \label{eq:decentralized_MPC}
\end{equation}
Although the decentralized MPC retains the mathematical structure of
the centralized formulation, decentralization changes three critical
components of the controller. First, the global continuum state \(\lambda(r,t)\) is replaced by a robot-dependent state reconstructed from
\(\mathcal N_i(t)\). Second, the global MIPS target \(M^*(t)\) is converted into a detection-radius-dependent local target \(M_D^*(t)\). Third, because jointly solving a distributed controller with all neighbors requires significant communication and computational power, defeating the purpose of decentralization, a local prediction of neighbor behavior is introduced to forecast the neighborhood dynamics without joint optimization.

At each MPC update time \(t_0\), robot \(i\) reconstructs the required
local continuum state using Gaussian kernels centered only at the robots
in \(\mathcal N_i(t_0)\). Depending on the selected prediction model,
this produces the local orientation distribution \(\psi_i\), the local
density-polarization state \((\lambda_i,p_i)\), or the local density \(\lambda_i\). Robots outside \(\mathcal N_i(t_0)\) do not contribute to robot \(i\)'s state estimate or the controller \eqref{eq:decentralized_MPC}.

The local target is not directly equal to the desired global MIPS
index because a MIPS index computed from a truncated observation region
has a detection-radius-dependent bias. For each \(D\), an empirical
calibration step is performed over varying swarm distributions to identify the mapping between \(M^*\) and \(M_D^*\).
Thus, \(M_D^*(t)\) represents the mean local MIPS level
whose ensemble average corresponds empirically to the desired global
target \(M^*(t)\). Thus, global target matching emerges automatically if every agent achieves \(M_D^*(t)\). 

Jointly predicting the optimal inputs with all neighboring robots would require exchanging candidate control trajectories or performing a distributed optimization at every MPC update. To close the
local prediction model without this coordination, robot \(i\) assumes
that all robots currently in \(\mathcal N_i(t_0)\) apply its candidate
input \(u_i(t)\) over the prediction horizon:
\(
u_j(\tau)=u_i(\tau),
\
j\in\mathcal N_i(t_0),
\
\tau\in[t_0,t_0+T].
\)
This shared-local-input assumption is motivated by the overlap of
neighboring observation regions: nearby robots possess similar
local density states, and are thereby expected to select similar actions under the same controller. This assumption is not required to hold at all times, and is perhaps valid only for physically close neighbors. However, its prediction error is eliminated at the next MPC update when the neighborhood and local density state are reconstructed from new measurements.

\section{Simulation Results}
\label{sec: sim results}
% \begin{figure}
%     \centering
%     \includegraphics[width=0.65\linewidth]{Figures/mips_random_initial_setup.png}
%     \caption{Simulation setup: starting from random initial conditions, \(500\)~robots are moving on a \(65\)~mm radius circular shaker plate, with the outer \(20\)~mm yellow slope to guide robots back towards the center flat area.}
%     \label{fig:setup}
% \end{figure}

To observe the behavior of the proposed controllers, \eqref{eq: centralized MPC} and \eqref{eq:decentralized_MPC}, we performed simulations using parameters extrapolated from the real BristleBots experiment data in \cite{MIPS_index}. 
Each simulation initializes \(500\) robots of radius \(1.5\)~mm at random, non-overlapping positions within a circular arena comprising a flat region of radius \(65\)~mm and a \(20\)-mm sloped boundary that redirects robots toward the center.
Our proposed MPC controllers are tested against the classic bang-bang controller, with vibration modes leading to aggregation and diffusion. Both centralized and decentralized controllers are tested against a step function and then a sine function as MIPS target \(M^*(t)\).

\begin{figure}
    \centering
    \includegraphics[width=0.99\linewidth]{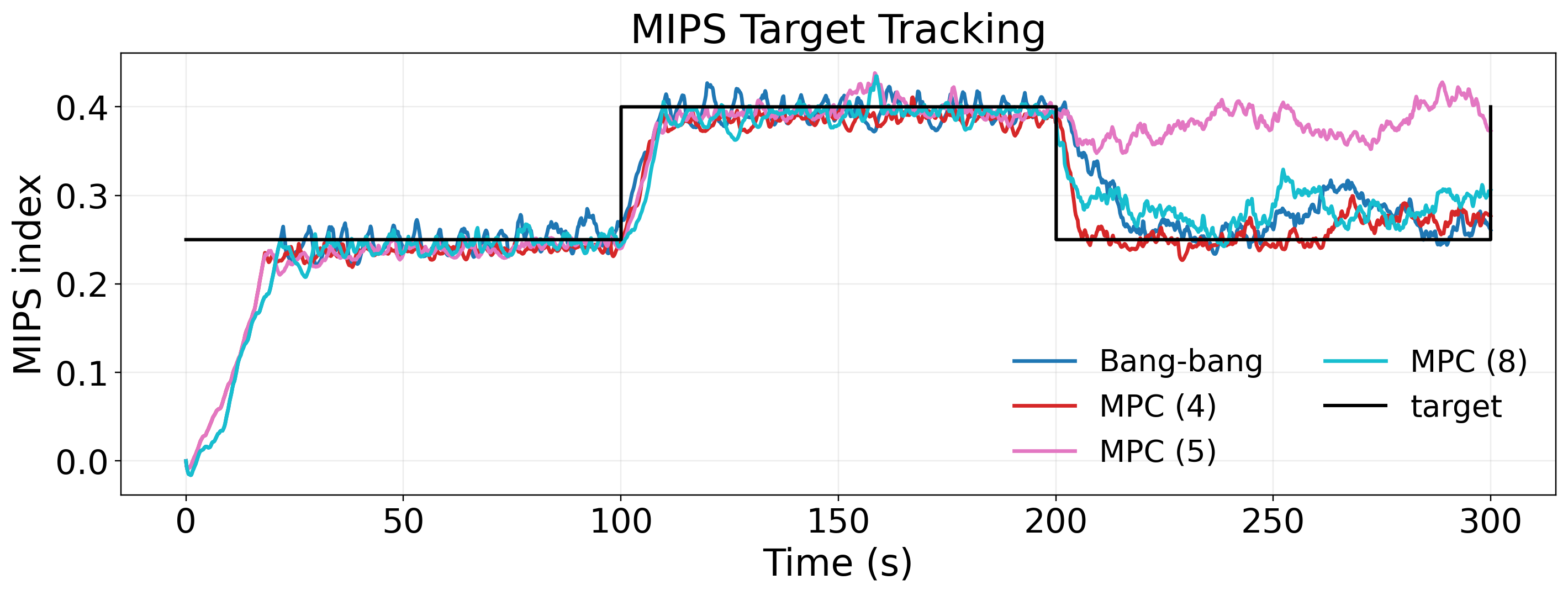}\\
    \includegraphics[width=0.31\linewidth]{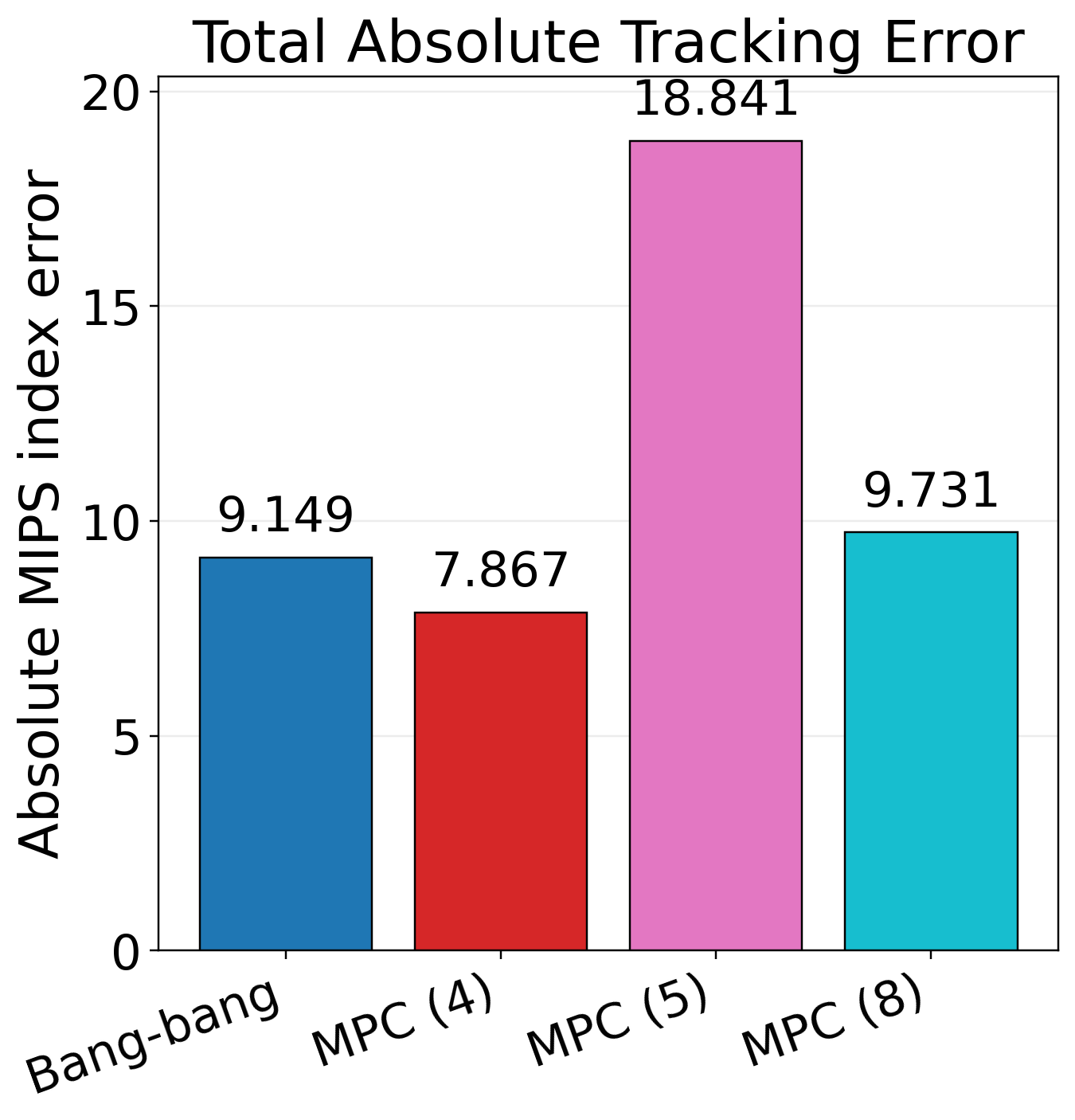}
    \includegraphics[width=0.32\linewidth]{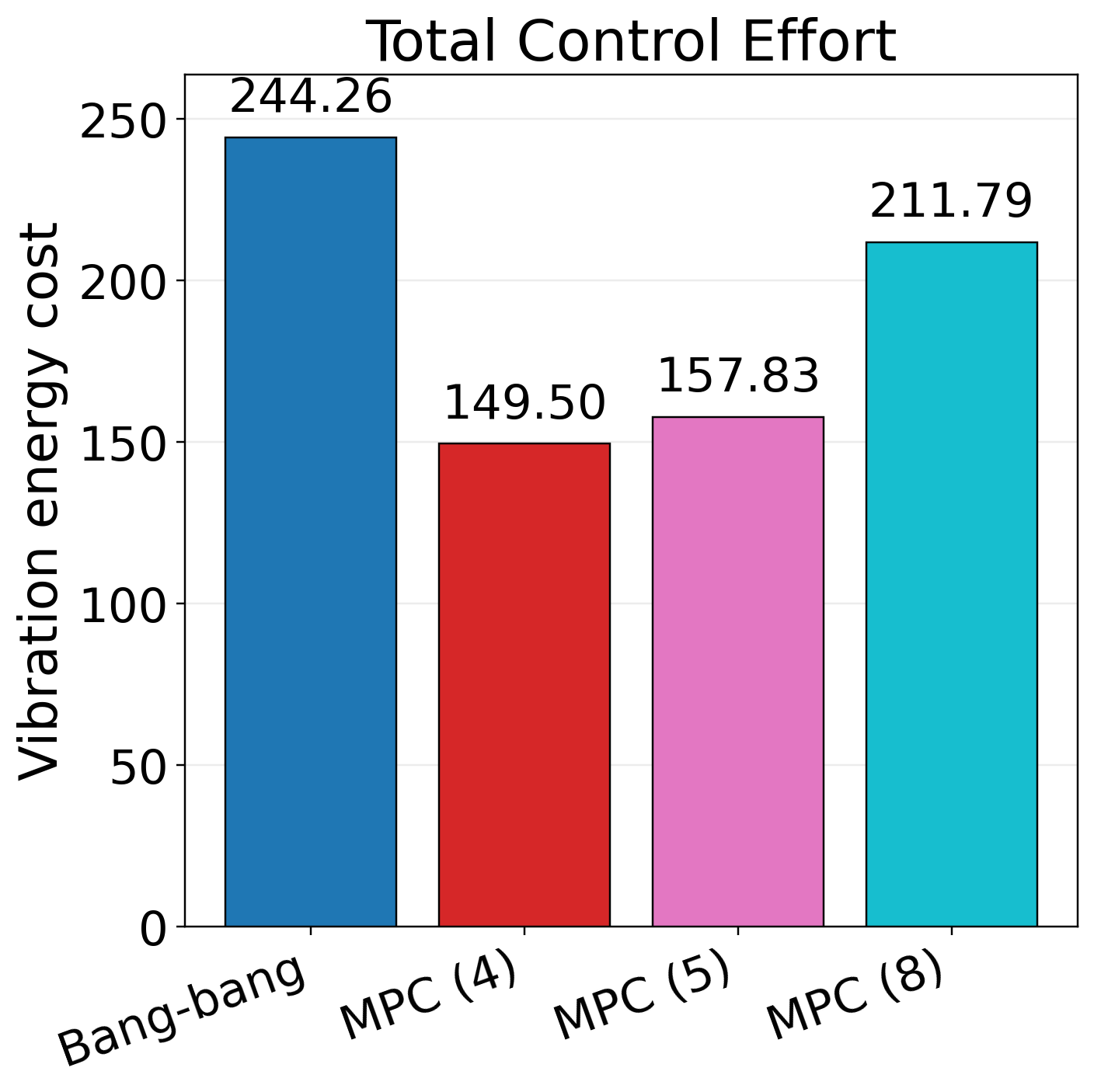}    \includegraphics[width=0.34\linewidth]{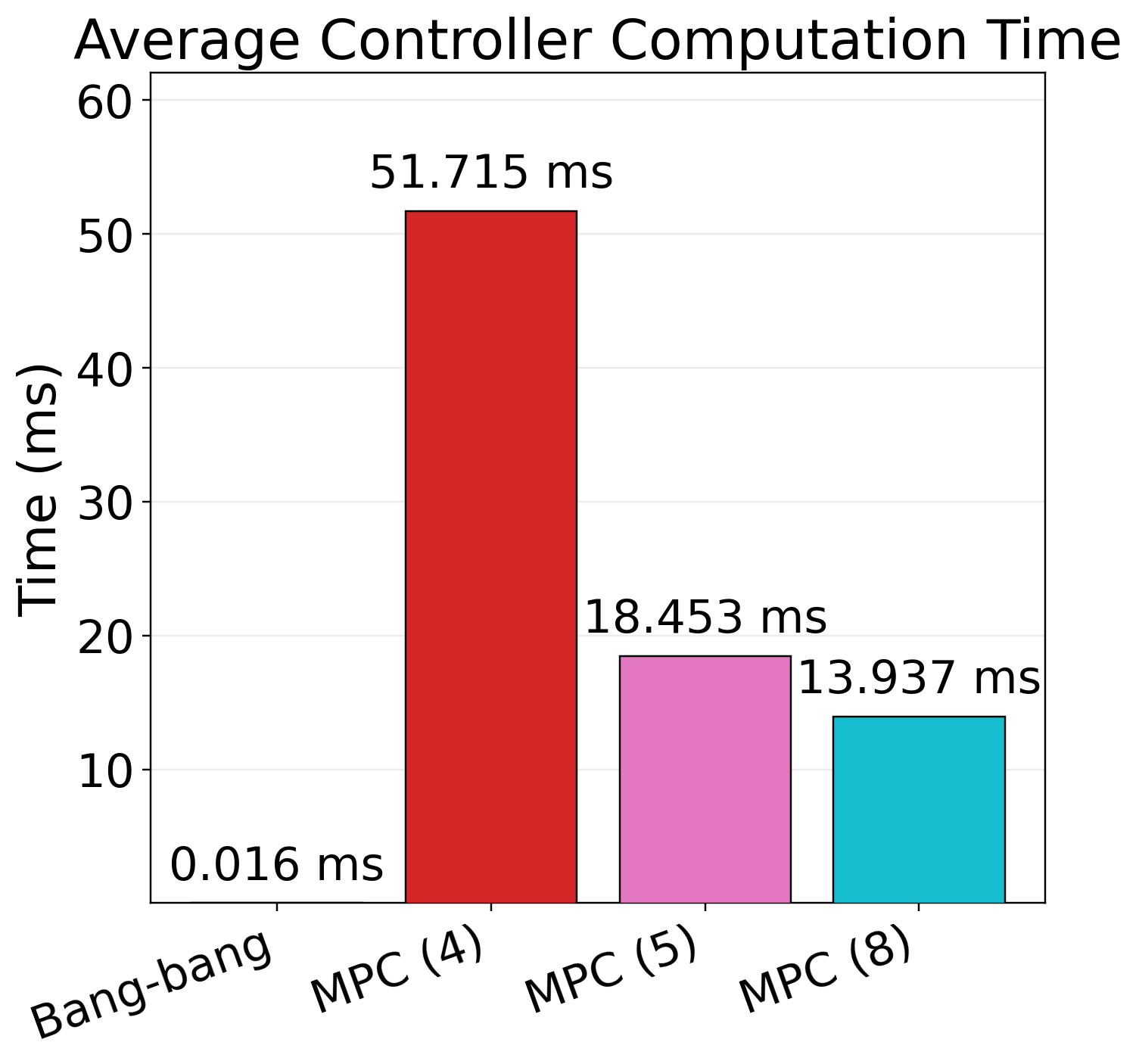}
    
    \caption{Centralized step-reference results: global MIPS-index tracking, accumulated absolute tracking error, normalized actuation effort, and average computation
    time per control update.}
    \label{fig:centralize_step}
\end{figure}

\subsection{Centralized Testing}
% The results of the step function test are shown in Fig.~\ref{fig:centralize_step}. As the MIPS index plot shows, the bang-bang controller has more oscillation when tracking a constant target value compared to the MPCs. Out of the three MPCs, the full FP model in~\eqref{eq: fp_model} is able to track the reference signal yielding the smallest tracking error, thus outperforming the bang-bang controller. This is the expected behavior, as this model retains most information with its explicit heading \(\theta\) state. 
% As a trade-off, the FP model has the longest computation time per step, with only a marginal advantage over the approximated models. 
% In terms of the control effort, all MPC approaches are cheaper than the bang-bang controller, with the FP-based one being the overall best, yielding better tracking and less energy spent. 

The simulation results are presented in Fig.~\ref{fig:centralize_step},
which compares the three MPC formulations with the bang-bang baseline.
The tracking plot in the upper panel shows that all four controllers
respond rapidly to the initial reference and successfully track the
increase in the target aggregation level. The bang-bang controller (blue)
exhibits greater oscillation around each constant target, whereas the
MPCs generally produce smoother responses by accounting for the
predicted density evolution over the control horizon. However, following
the dilution command at \(200\)~s, the density-polarization controller (magenta)
fails to maintain target tracking for the remainder of the simulation.
During this transition, the full FP controller (red) based on
\eqref{eq: fp_model} responds noticeably faster than the other
controllers. This is the expected behavior, as this model retains most information with its explicit heading \(\theta\) state.

The bar plots at the bottom of Fig.~\ref{fig:centralize_step} quantify these performance differences. The full
FP-based controller achieves the lowest accumulated absolute
tracking error, whereas the density-polarization controller produces
the highest error as it fails to track during the final
\(100\)~s. The bang-bang controller also marginally outperforms the
density-only model in terms of tracking error. However, its aggressive
switching between the two actuation modes results in the highest total
control effort. Because the MPC formulations explicitly penalize
actuation effort in their objectives, all three MPCs spent less control
effort than the bang-bang controller, with the full FP
controller achieving the lowest value. The final panel reports the
average computation time per control update. As expected, propagating
the complete position and orientation information makes the full
FP controller more than twice as computationally expensive
as the two MPCs. The polarization model spent more computational resources since it involves two coupled PDEs rather than a single field in the Density-Flux model.
Finally, the rule-based mode selection of the bang-bang controller incurred negligible computational cost.

\begin{figure}
    \centering
    \includegraphics[width=0.99\linewidth]{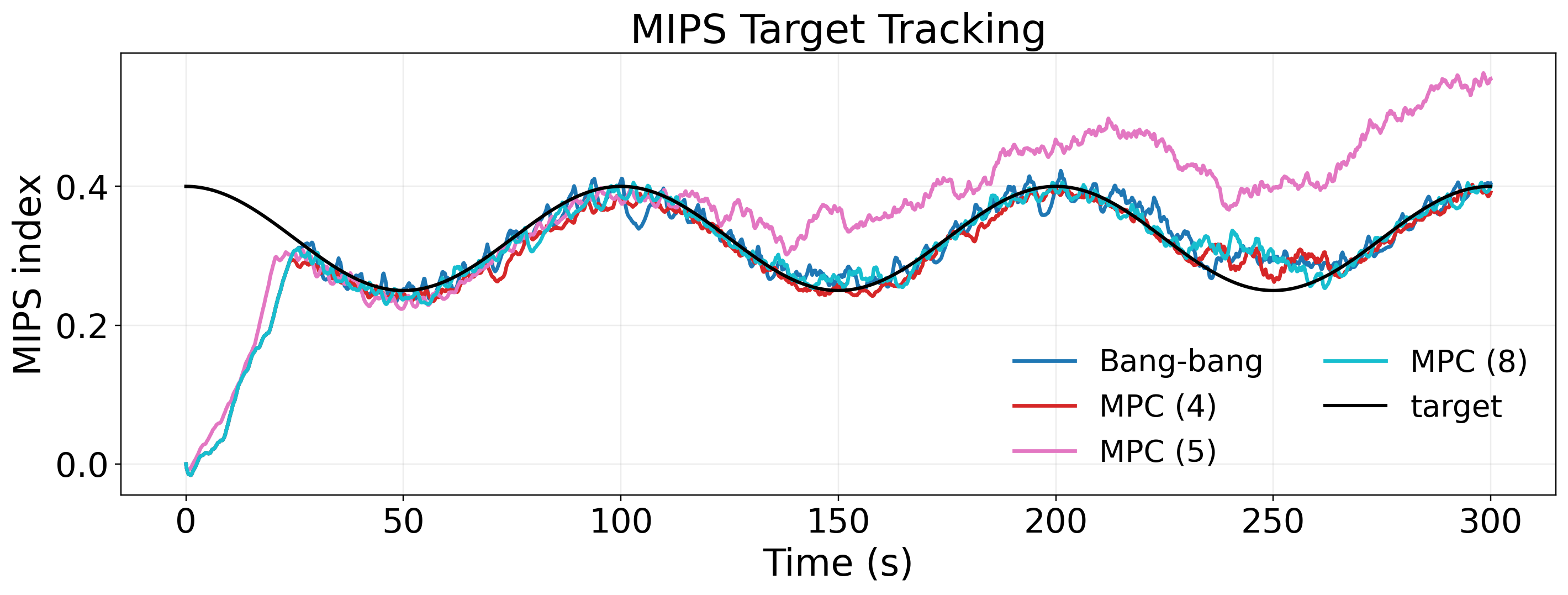}\\
    \includegraphics[width=0.31\linewidth]{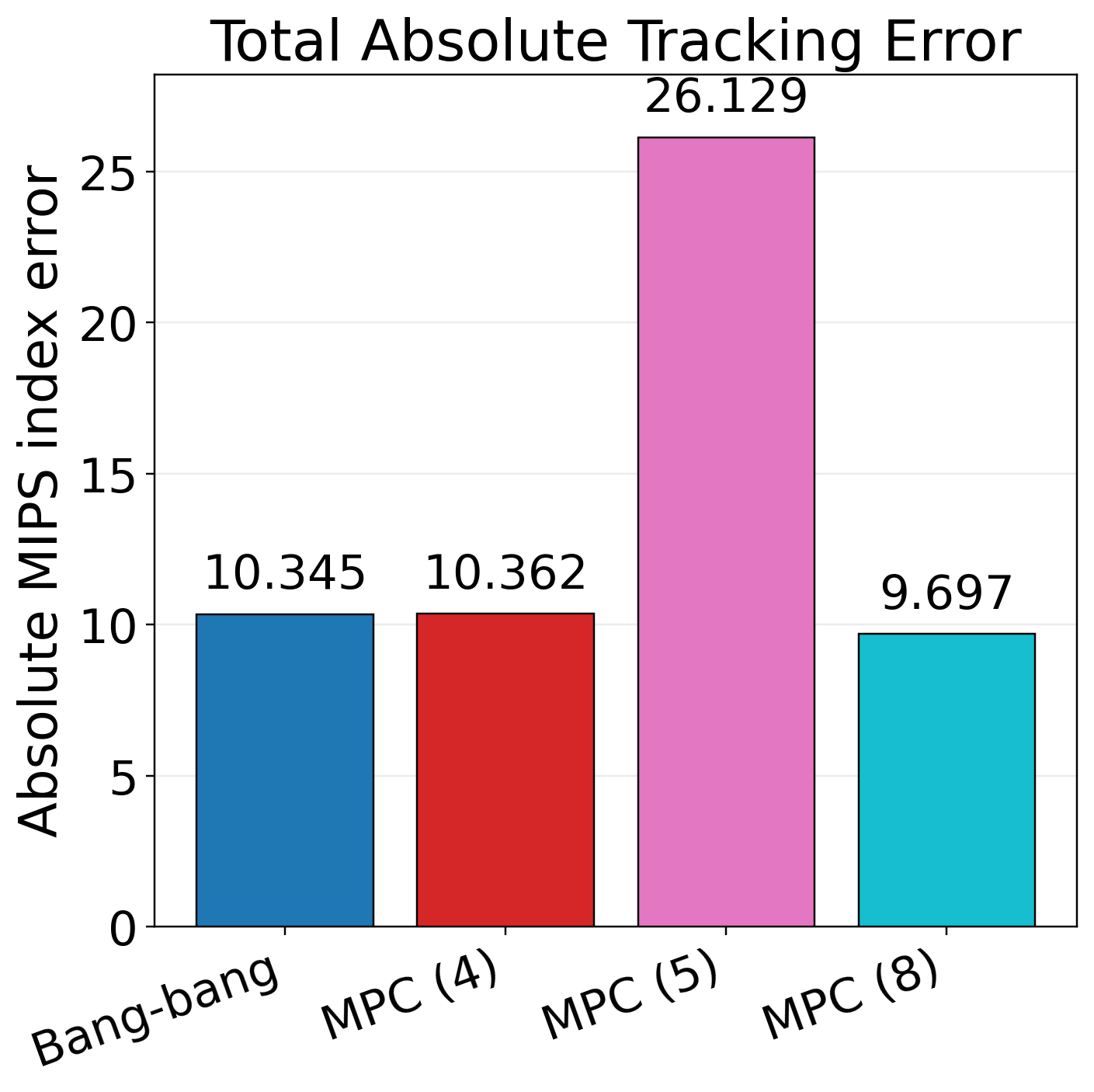}
    \includegraphics[width=0.32\linewidth]{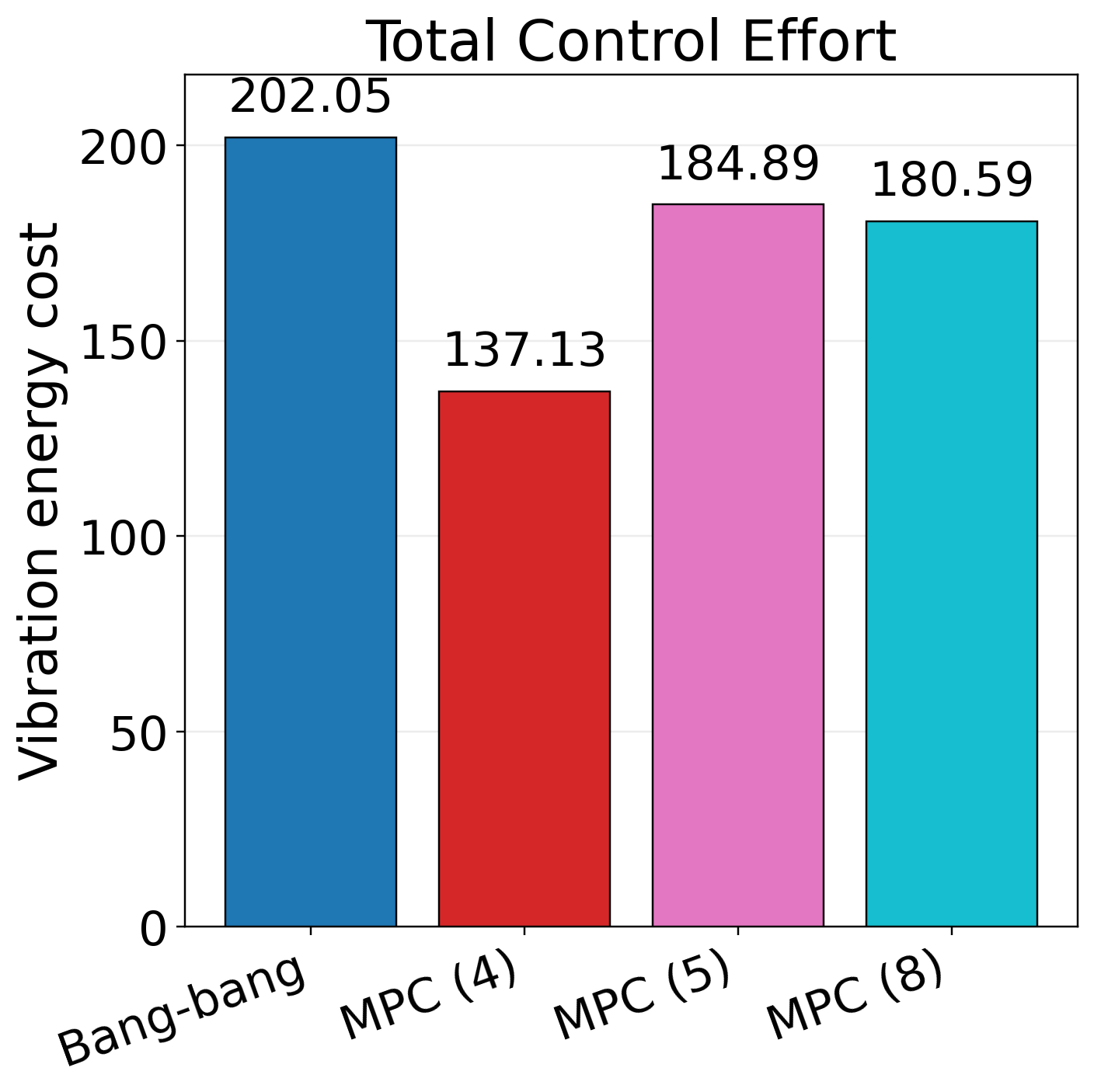}    \includegraphics[width=0.34\linewidth]{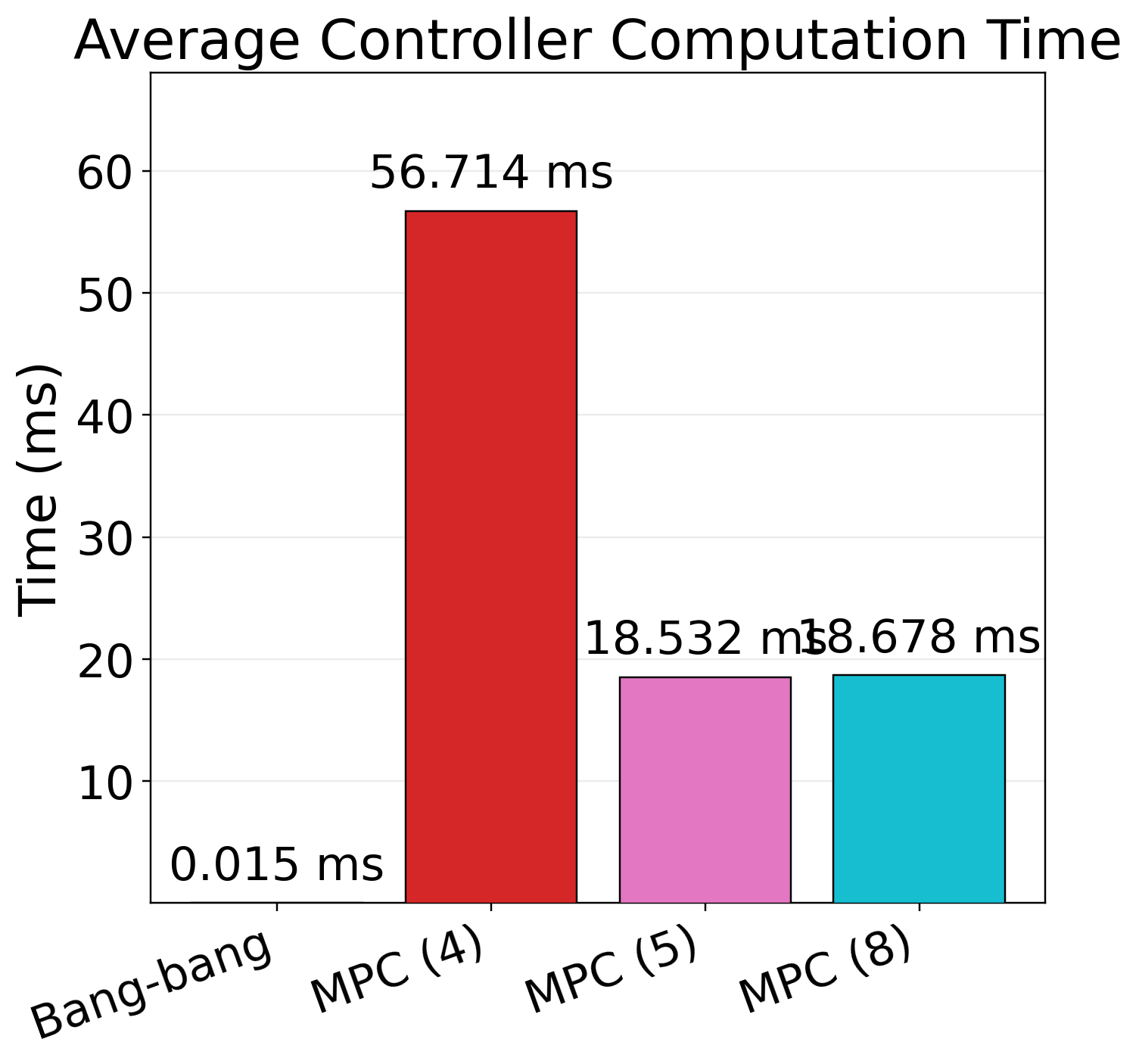}
    
    \caption{Centralized sinusoidal-reference results: global MIPS-index tracking, accumulated absolute tracking error, normalized actuation effort, and average computation
    time per control update.}
    \label{fig:centralize_sine}
\end{figure}

Next, we evaluated the controllers using a sinusoidal MIPS reference, as shown in Fig.~\ref{fig:centralize_sine}. The bang-bang controller exhibits significant oscillations, yet it is able to track the increasing and decreasing trends of the reference MIPS index. The density-only MPC (cyan) achieves the best tracking performance, followed by the full FP (red) controller. The density-polarization controller (magenta), instead, loses the reference after approximately \(100\)~s. %The improved relative performance of the bang-bang controller for the sinusoidal reference can be explained by its switching behavior. When the target increases, the controller remains in the aggregation mode for a longer interval, whereas a decreasing target produces a longer interval in the diffusion mode. These prolonged intervals reduce the rapid switching observed when regulating a constant reference. 
This behavior suggests that retaining only the first angular moment is insufficient, since distinct heading distributions can have the same mean polarization while producing different subsequent density dynamics.
Among the remaining MPCs, the full FP controller requires the least total energy expenditure, while the computation-time results follow the same trend as in the step-reference test, with the full FP model incurring the highest computational cost.

\begin{remark}[Effect of prediction horizon]
Longer horizons degraded performance despite remaining numerically
stable, likely because the continuum predictors omit finite robot
size and maximum packing, causing prediction errors to accumulate
over the horizon.
\end{remark}

Together, the step and sinusoidal tests indicate that approximating the
angular harmonic components with only the first-order polarization term can oversimplify the swarm dynamics. A higher-fidelity moment model would require retaining the second-order, and potentially additional, angular harmonics, resulting in a larger coupled PDE system. The full
FP and density-only MPCs provide more favorable overall
performance, although neither uniformly dominates the bang-bang
controller across every metric. Their clear advantages are smoother
tracking responses and lower actuation waste, obtained at the cost of
additional computation. In the following section, these controllers are
extended to the decentralized setting, with the decentralized bang-bang
controller retained as the baseline.

\subsection{Decentralized Testing}
To examine the role of local information in the decentralized controller \eqref{eq:decentralized_MPC}, we first characterize two consequences of restricting detection radius \(D\): (i) the relationship between local and global MIPS measurements and (ii) the effect of \(D\) on closed-loop tracking performance.

\begin{figure}
    \centering
    \includegraphics[width=0.75\linewidth]{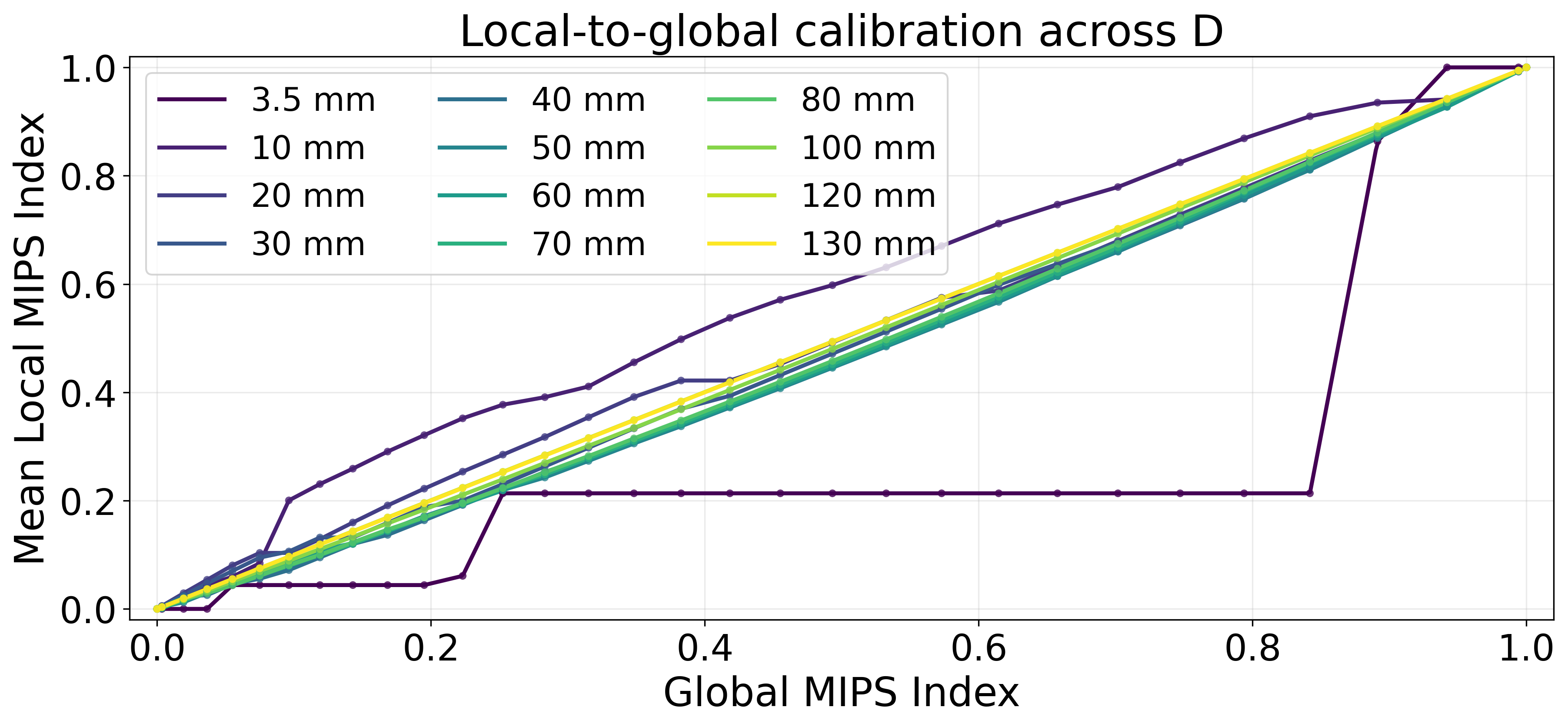}\\
    \includegraphics[width=0.75\linewidth]{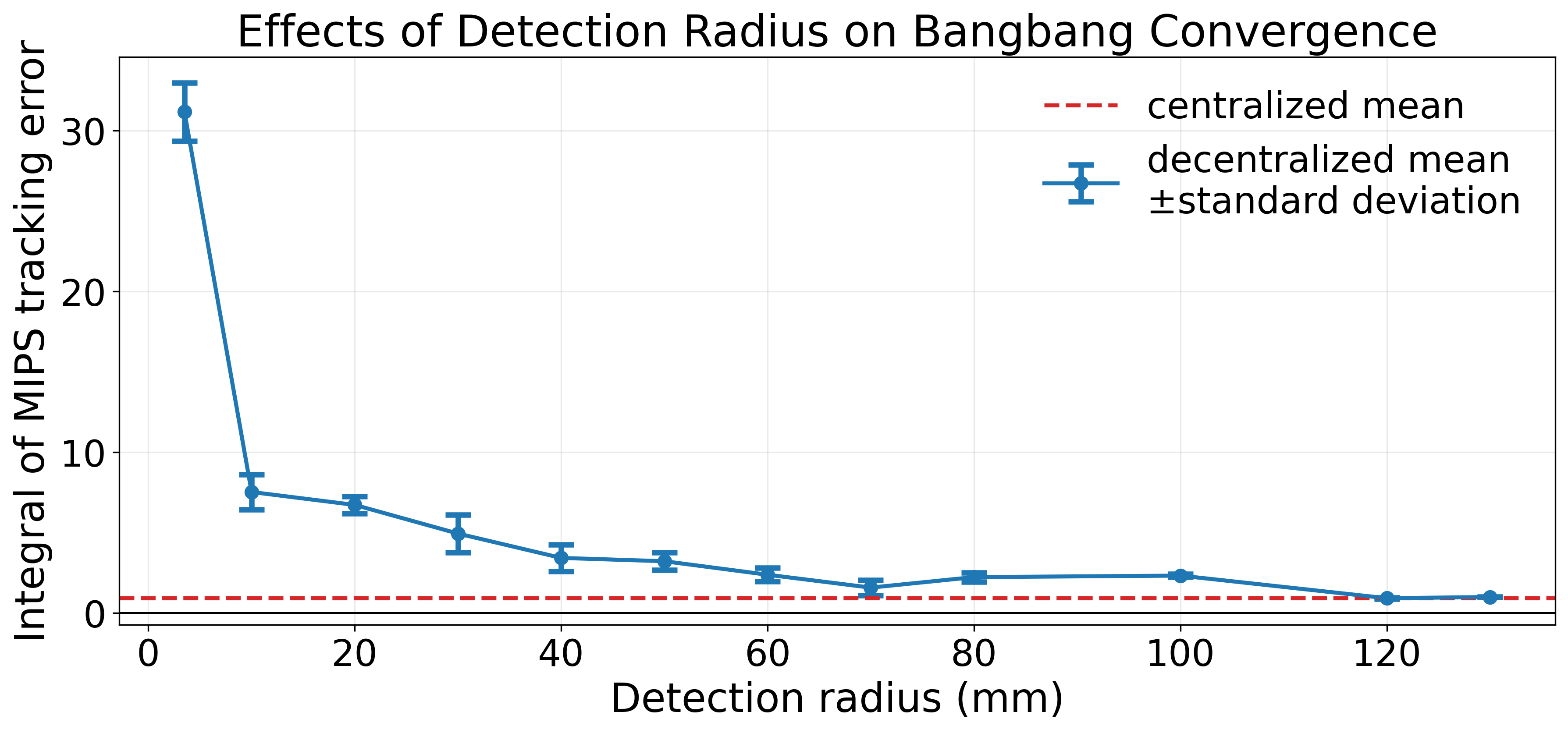}
    \caption{Effect of detection radius: (a) relationship between mean local and global MIPS indices; (b) mean time-integrated absolute tracking error over the final \(20\)~s, with standard-deviation error bars and the centralized benchmark.}
    \label{fig: detection radius}
\end{figure}

First, recall that the MIPS index computed from a truncated observation generally does not equal the global aggregation level. Let
\(M_{D,i}\) denote the local MIPS index measured by robot \(i\),
We empirically characterize the relationship between \(M\),
the global MIPS index, and \(\frac{1}{N} \sum_{i = 0}^N M_{D,i}\), the local MIPS mean, using varying simulated swarm distributions. The resulting measurements are shown at the top of Fig.~\ref{fig: detection radius} and are subsequently used to
construct the detection-radius-dependent target map:
\(g_D: M_D^*=g_D(M^*).\)

% \begin{figure}
%     \centering
%     \includegraphics[width=1.0\linewidth]{Figures/decentralize/all_local_vs_global_mips_kde.png}
%     \caption{Relationship between mean local and global MIPS values under different detection radii.}
%     \label{fig:mips_local_global_switch}
% \end{figure}

The global and local calculations use the same density construction,
variance evaluation, and normalization procedure. Their only difference
is the set of robots contributing to the density estimate: the global
measurement uses the complete swarm, whereas \(M_{D,i}\) uses only the
robots in \(\mathcal N_i\). Consequently, if \(D\) is at least the
diameter of the entire domain, \(D\geq 2(65)=130\)~mm, every
robot observes the complete swarm and the two measurements should
coincide, as confirmed in Fig.~\ref{fig: detection radius}. 

Conversely, the relationship becomes weaker and more variable as the
detection radius decreases. At \(D=3.5\)~mm, corresponding
approximately to the collision-sensing case discussed in
Remark~\ref{remark:Collsion_based_decentral}, the local measurements
contain insufficient spatial information to reliably infer the global
aggregation level. The empirical calibration \(g_D\) can compensate for
systematic differences in scale and bias, but it cannot recover global
information that is absent from such a small local neighborhood.

The detection radius also affects the temporal performance of the
closed-loop system. Smaller neighborhoods provide more variable density estimates, causing nearby robots to construct
different local MIPS values and potentially select different actuation inputs. To quantify this effect, the bottom plot of Fig.~\ref{fig: detection radius} reports the mean and standard deviation of the time-integrated absolute
tracking error over the final \(20\)~s of \(20\) independent
\(100\)-s decentralized bang-bang simulations,
\(E_D
=
\int_{80}^{100}
\left|M(t)-M^*(t)\right|\,dt\), 
which is sufficient since the centralized convergence time is only \(20\)-\(30\)~s.
The results show that both the tracking error and its standard deviation generally increase as \(D\) decreases. Collision-range
sensing produces the largest degradation, whereas the result at
\(D=130\)~m approaches the centralized benchmark because each
robot effectively observes the complete swarm.

\begin{figure}
    \centering
    \includegraphics[width=0.99\linewidth]{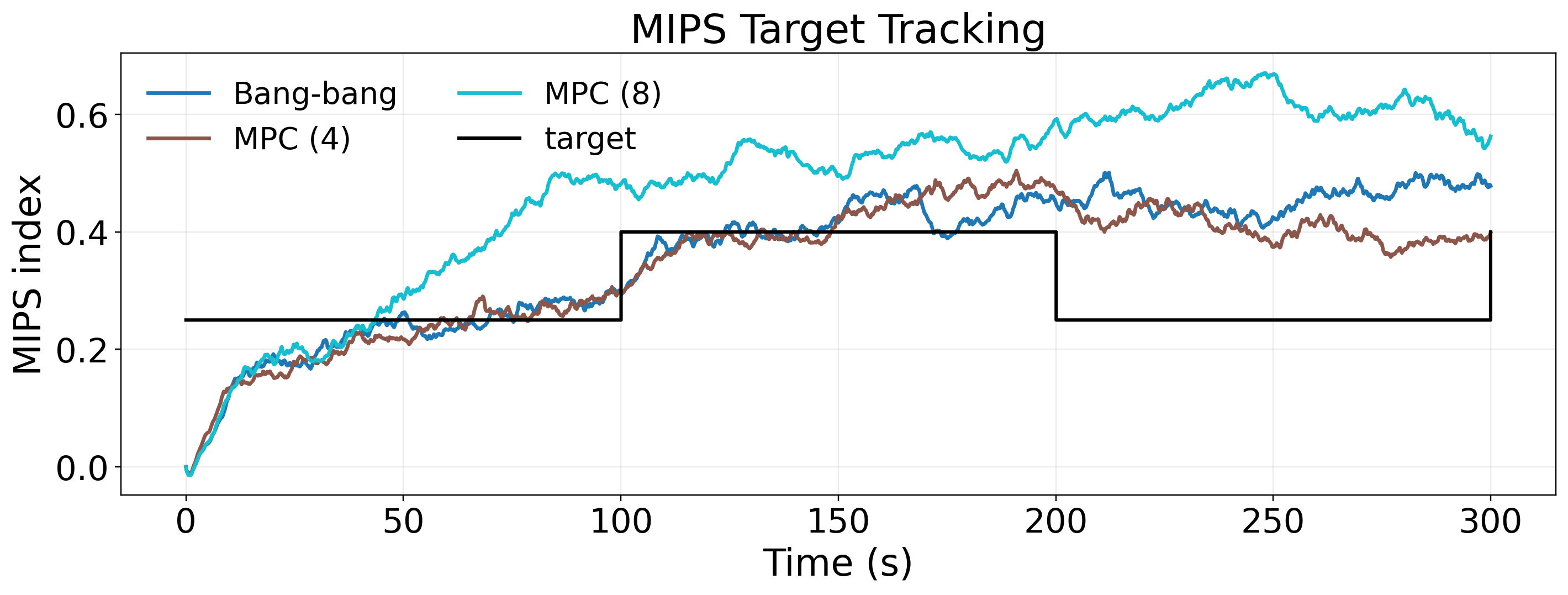}\\
    \includegraphics[width=0.31\linewidth]{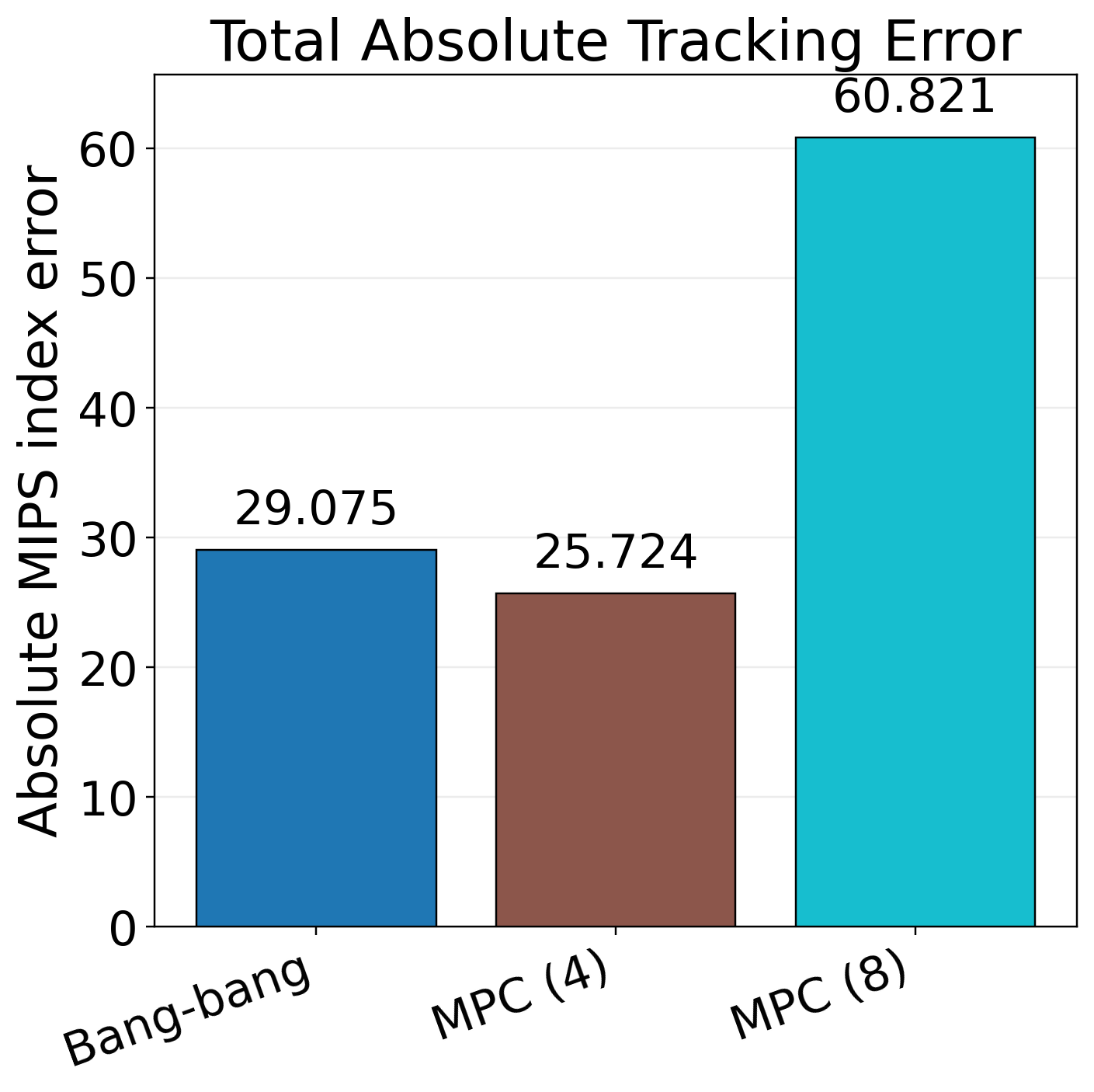}
    \includegraphics[width=0.32\linewidth]{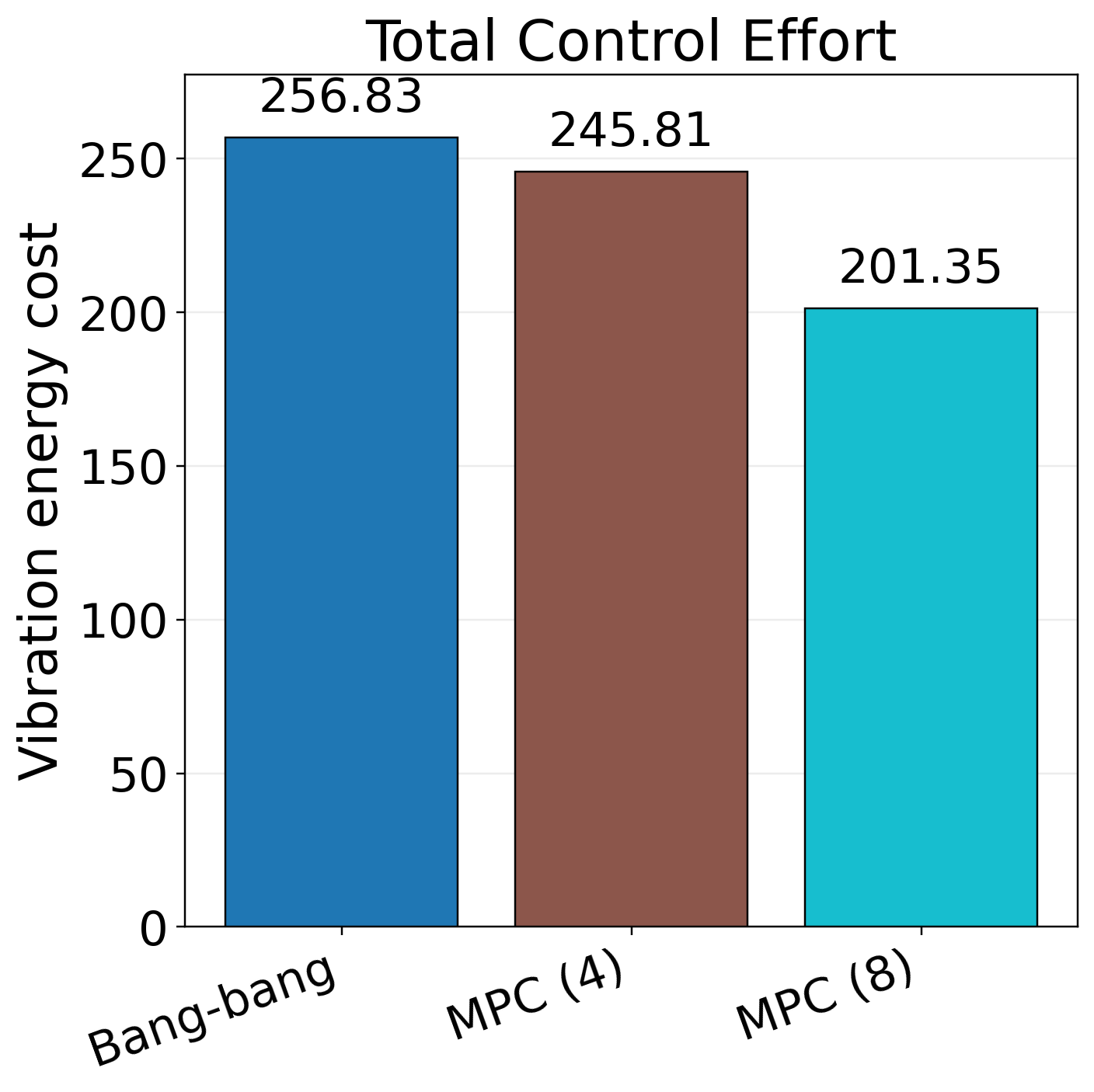}    \includegraphics[width=0.34\linewidth]{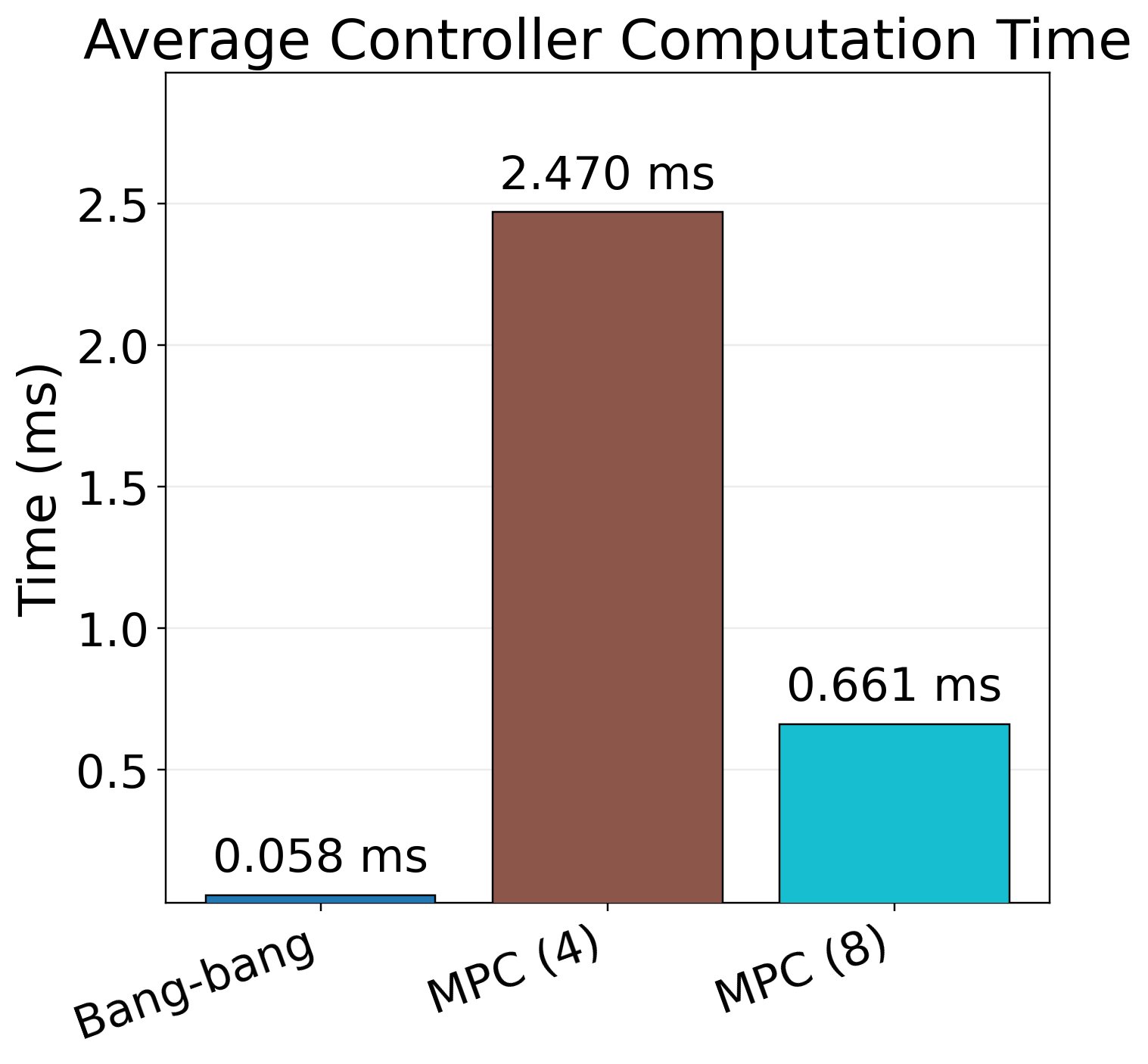}
    
    \caption{Decentralized step-reference results for
    \(D=40\)~mm: global MIPS-index tracking, accumulated absolute tracking error, normalized actuation effort, and average computation
    time per control update.}
    \label{fig:decentralize_step}
\end{figure}

Having quantified the effect of the detection radius, we next evaluate
the decentralized bang-bang and MPC controllers using
\(D=40\)~mm. This intermediate radius provides a relatively small detection radius while maintaining a sufficiently
informative relationship between the local and global MIPS indices. The decentralized step-reference results are presented in Fig.~\ref{fig:decentralize_step}. Initially, the calibrated local targets \(M_D^*\) exceed the local measurements \(M_{D,i}\), causing the three decentralized controllers to aggressively drive the swarm toward aggregation. The rate of increase of the global MIPS index subsequently decreases as the local measurements approach \(M_D^*\), briefly plateauing at \(120\)~s. At \(200\)~s, the reference steps to a lower aggregation level, thereby commanding dispersion. None of the controllers produces a sufficiently rapid decrease in the global MIPS index during the remaining simulation time.
Nevertheless, the trajectory produced by the full FP and Density-Flux controllers begins to exhibit a slight decrease near the end of the \(300\)-s simulation.

This delayed response is consistent with the detection-radius study in Fig.~\ref{fig: detection radius}. In those simulations, the decentralized bang-bang controller required approximately two to three times longer than its centralized counterpart to settle near a constant target. Thus, although none of the decentralized controllers achieved accurate tracking after the reference change, the late decrease produced by the two MPCs suggests that their response to the dispersion command is delayed on a substantially longer timescale under limited local information.
The quantitative results exhibit trends similar to those observed in the centralized tests. 
%However, the differences in total actuation effort are smaller in the decentralized case, as each robot optimizes its input using only a local prediction of its neighbors' actions and cannot account directly for the resulting swarm-wide energy expenditure. This limited local information also produces slower responses to target fluctuations, as the robot cannot access the density information outside of its detection radius. Therefore, the swarm requires longer intervals to achieve actuation towards the target, thereby reducing the energy advantage of the FP controller over the bang-bang controller. Nevertheless, with its predictive power combined with the full angular state, it still follows the global target most closely, at the expense of computational delays. 

\begin{figure}
    \centering
    \includegraphics[width=0.99\linewidth]{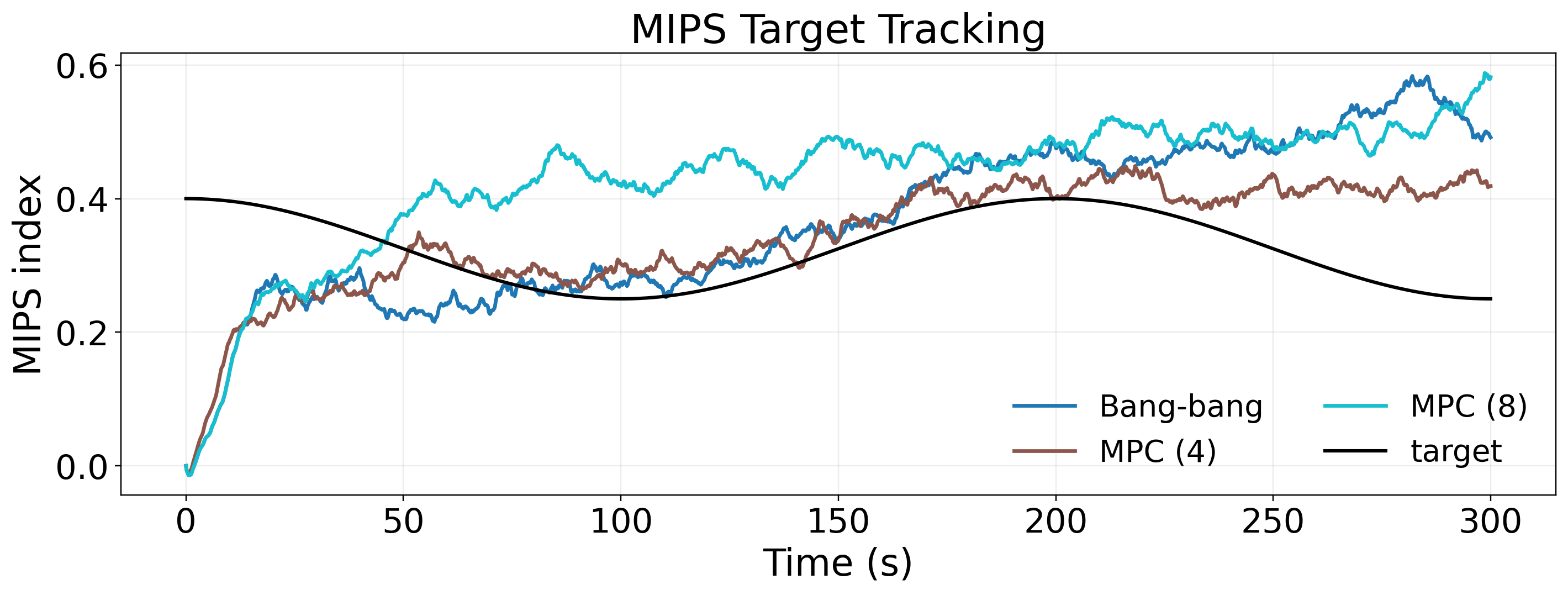}\\
    \includegraphics[width=0.31\linewidth]{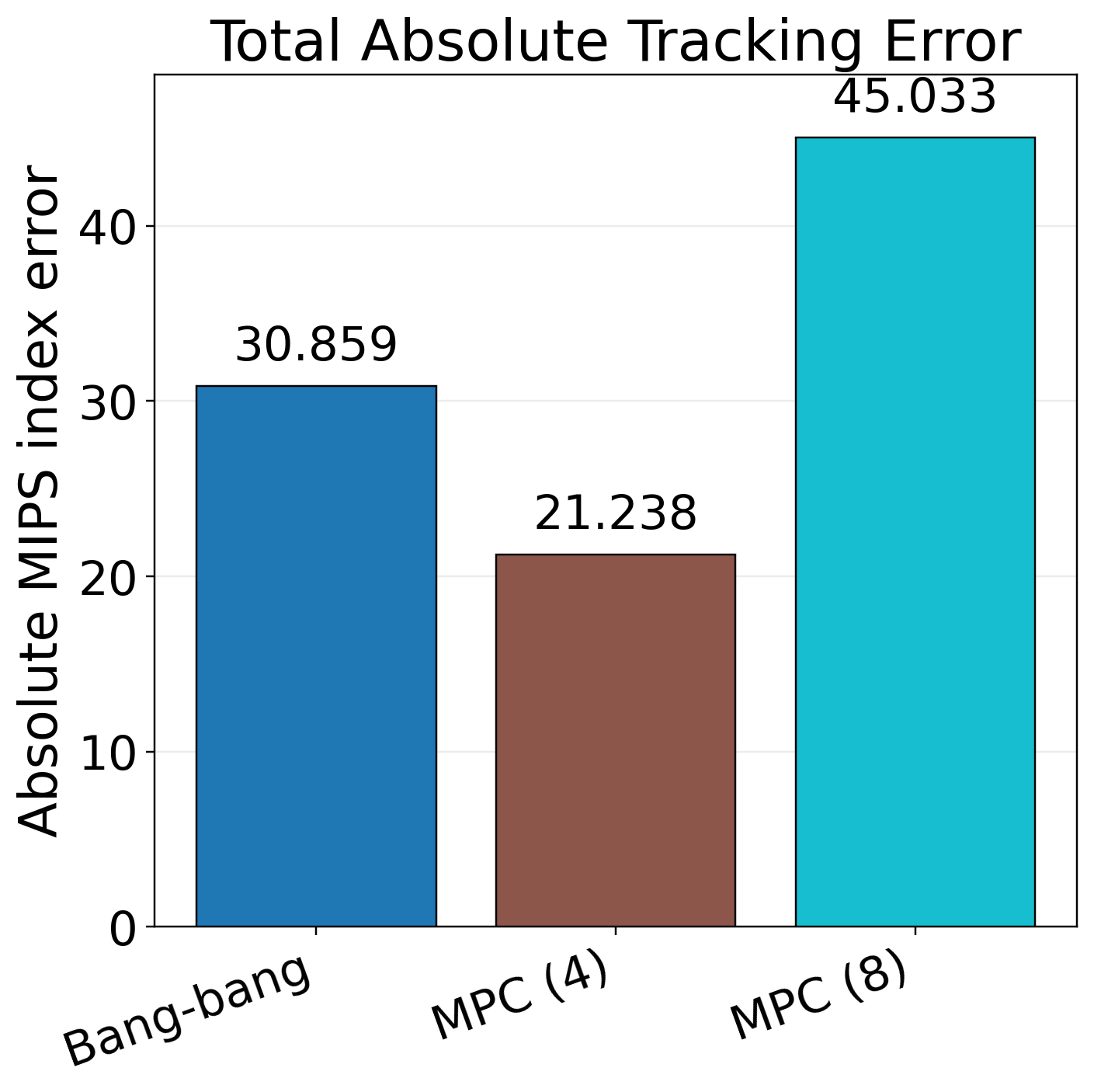}
    \includegraphics[width=0.32\linewidth]{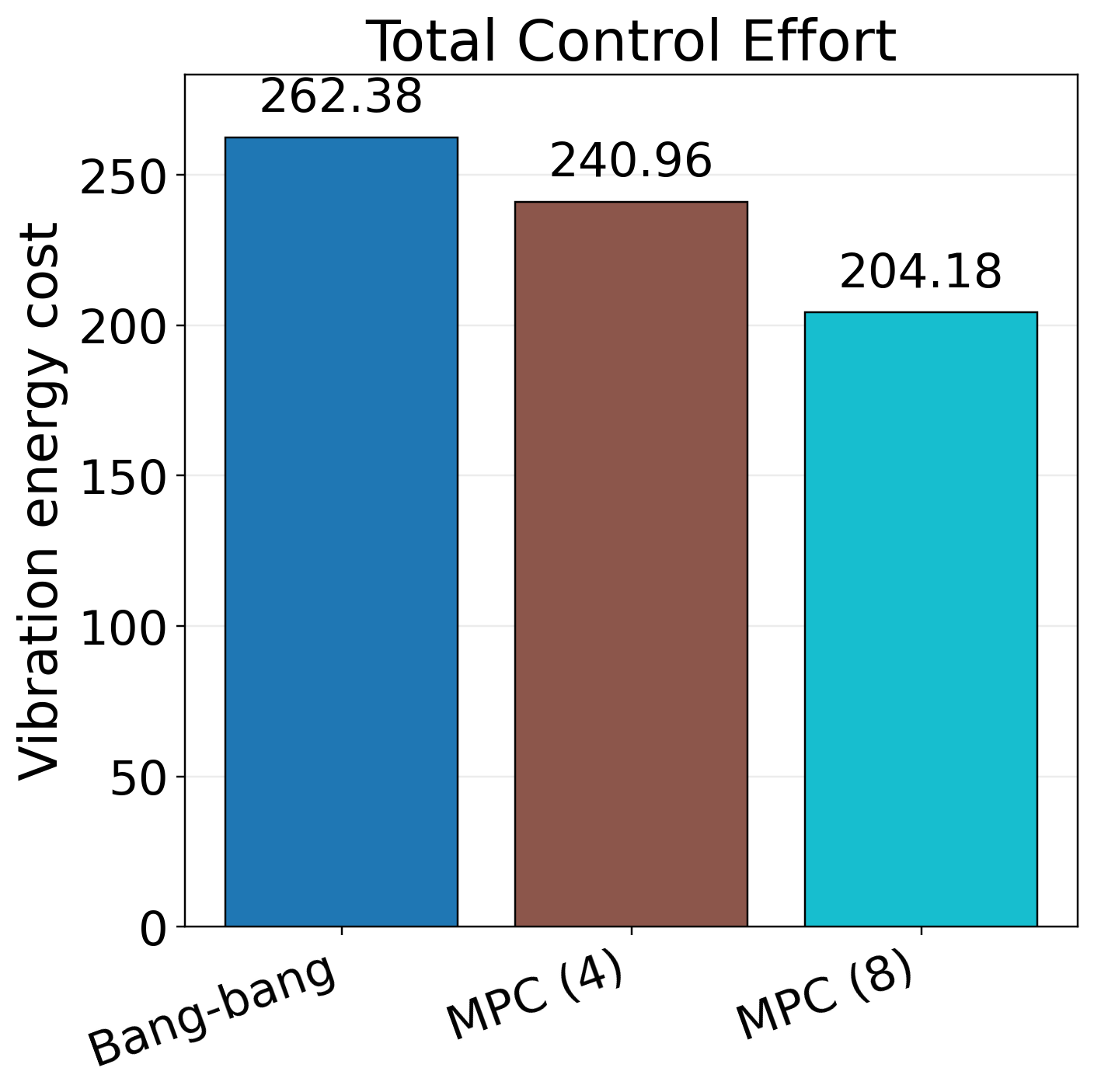}    \includegraphics[width=0.34\linewidth]{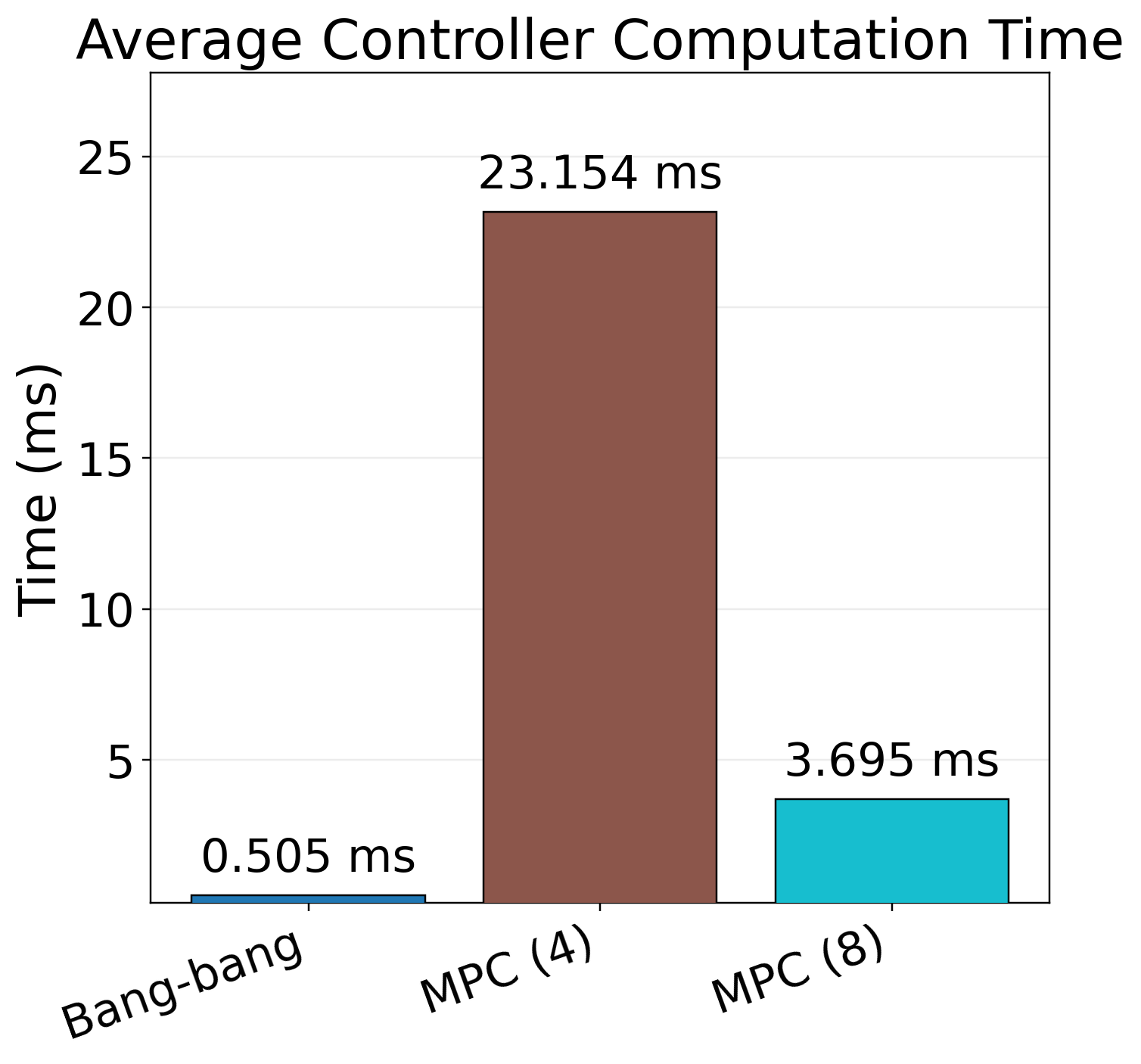}
    
    \caption{Decentralized sinusoidal-reference results for
    \(D=40\)~mm: global MIPS-index tracking, accumulated absolute
    tracking error, normalized actuation effort, and average computation
    time per control update.}
    \label{fig:decentralize_sino}
\end{figure}

For the sinusoidal reference experiment, the results are shown in Fig.~\ref{fig:decentralize_sino}. Both decentralized simulations exhibit similar trends. Specifically, the full-information FP model \eqref{eq: fp_model} outperforms the other controllers on every metric, at the cost of higher computation time. The density-flux model \eqref{sec: density_model} falls behind the bang-bang controller as a consequence of decentralization. This shows that the detection radius limitation on available information around a robot can negatively affect the predictions of \eqref{sec: density_model}, which relies on assumptions of the swarm parameters based on large particle models.

Therefore, the decentralized experiments show that the empirical local-to-global calibration %can partially correct systematic differences between local and global MIPS measurements, but it 
cannot recover spatial information outside the neighborhood. At \(D=40\)~mm, the controllers initiate the appropriate aggregation and dispersion behaviors, but the collective global response is substantially slower than in the centralized case. Consequently, the quality of predictive control depends not only on the selected continuum model, but also on whether the locally observed density contains sufficient information to predict changes in the global scale. These results highlight the role of space-time coupling: restriction in local (in space) observability degrades the predictive accuracy (in time) and closed-loop response of the PDE-based MPC. Improving this decentralized performance requires addressing a limitation of the current local-to-global calibration: cropping can make a small local cluster appear fully aggregated or a locally uniform neighborhood fully dispersed, while zero-padding introduces artificial variance. Future work will therefore investigate communication-assisted distributed MIPS estimation, such as using consensus to fuse local density statistics while retaining decentralized control.

\section{Conclusion}
This paper presented a PDE-constrained MPC framework for directly regulating the MIPS index of a vibration-driven robotic swarm. Three continuum predictors retaining progressively less orientation information were evaluated in centralized and decentralized settings. The centralized results demonstrate a trade-off between predictive fidelity and computational cost: the full Fokker--Planck model provides the strongest overall performance, reduced models can remain competitive, and retaining only the first angular moment is insufficient for some transitions. The decentralized results show that restricted spatial observations weaken local-to-global MIPS estimation and produce slower temporal responses. Thus, effective predictive MIPS control depends on both model fidelity and information availability. Future work will investigate communication-assisted distributed estimators, such as consensus-based fusion of local density statistics, to improve global MIPS estimation while retaining decentralized control.

% \section{Acknowledgment}
% ChatGPT was used as a coding assistant, primarily to support data visualization and debug certain functions.

\bibliographystyle{IEEEtran}
\bibliography{references_noURL}
\end{document}